\documentclass{iopart}

\usepackage{color,graphicx}
\usepackage{amssymb}
\usepackage{mathrsfs}
\usepackage{supertabular}
\usepackage{color,graphicx}
\usepackage[bookmarksopen]{hyperref}
\usepackage{cite}

\def  \bef  {\begin{figure}}
\def  \eef  {\end{figure}}
\def  \be   {\begin{equation}}
\def  \ee   {\end{equation}}
\def  \ba   {\begin{array}}
\def  \ea   {\end{array}}
\def  \bea  {\begin{eqnarray}}
\def  \eea  {\end{eqnarray}}
\def  \beq  {\begin{eqnarray}}
\def  \eeq  {\end{eqnarray}}
\def  \nn   {\nonumber}
\def  \bd   {\begin{displaymath}}
\def  \ed   {\end{displaymath}}
\def  \bse  {\begin{subequations}}
\def  \ese  {\end{subequations}}
\def  \bwt  {\begin{widetext}}
\def  \ewt  {\end{widetext}}

\def  \ba   {{\bf{a_1}}}

\begin{document}

\title[]{Systematic study of charmonium production in pp collisions at LHC energies}

\author{Biswarup Paul, Mahatsab Mandal, Pradip Roy and Sukalyan Chattapadhyay}

\address{High Energy Nuclear and Particle Physics Division, Saha Institute
of Nuclear Physics,
1/AF Bidhannagar, Kolkata-700 064, INDIA}
\ead{biswarup.paul@saha.ac.in,  mahatsab@gmail.com}

\begin{abstract}
We have performed a systematic study of $J/\psi$ and $\psi(2S)$ production in $p-p$ collisions at different LHC energies and at different rapidities using the leading order (LO) non-relativistic QCD (NRQCD) model of heavy quarkonium production. We have included the contributions from $\chi_{cJ}$ ($J$ = 0, 1, 2) and $\psi(2S)$ decays to $J/\psi$. The calculated values have been compared with the available data from the four experiments at LHC namely, ALICE, ATLAS, CMS and LHCb. In case of ALICE, inclusive $J/\psi$ and $\psi(2S)$ cross-sections have been calculated by including the feed-down from $B$ meson using Fixed-Order Next-to-Leading Logarithm (FONLL) formalism. It is found that all the experimental cross-sections are well reproduced for $p_T >$ 4 GeV within the theoretical uncertainties arising due to the choice of the factorization scale. We also predict the transverse momentum distributions of $J/\psi$ and $\psi(2S)$ both for the direct and feed-down processes at the upcoming LHC energies of $\sqrt{s} =$ 5.1 TeV and 13 TeV for the year 2015.

\end{abstract}

\maketitle
\section{Introduction}
The understanding of the production of the heavy quarkonium (bound states of heavy quark ($Q$) and heavy anti-quark ($\bar Q$)) has been a long-term effort both experimentally and theoretically. The different treatments of the non-perturbative evolution of the $Q \bar Q$ pair into a quarkonium lead to various theoretical models. There are mainly three models widely used to describe the production of quarkonium: the Color Singlet Model (CSM), the Color Evaporation Model (CEM) and the NRQCD framework. 

The CSM, proposed right after the discovery of the $J/\psi$, assumes that the $Q \bar Q$ pair evolving in a quarkonium state is in a color-singlet (CS) state and the quantum numbers such as spin and angular momentum, are conserved after the formation of the quarkonium. The only inputs required in the model are the absolute value of the colour singlet $Q \bar Q$ wave function and its derivatives that can be determined from data of decay processes. Once these quantities are provided, the CSM has no free parameters~\cite{npb172}. The CSM at leading-order, predicts well the quarkonium production rates at relatively low energy~\cite{arx94}, but fails to describe the data for charmonium measured by CDF experiment in $p-\bar p$ collisions~\cite{prl69} probably because it ignores the 
fragmentation processes from higher states or $B$ mesons, dominant at Tevatron energies~\cite{prl71}. Recently it has been revived, with the computation at higher orders in the strong coupling constant $\alpha_{s}$ expansion~\cite{plb653,prl98,prl101}, since it was found to better accomodate polarization results from Tevatron with respect to NRQCD.

The CEM~\cite{arx041} is a phenomenologically successful model and was first proposed in 1977~\cite{plb67}. In the CEM, the cross-section for a quarkonium state $H$ is a fraction $F_{H}$ of the cross-section of the produced $Q \bar Q$ pairs with invariant mass below the $M \bar M$ threshold, where $M$ is the lowest mass meson containing the heavy quark $Q$. This cross-section has an upper limit on the $Q \bar Q$ pair mass but no constraints on the colour or spin of the final state. The $Q \bar Q$ pair is assumed to neutralize its colour by interaction with the collision-induced colour field by "colour evaporation". An important feature is that the fractions $F_{H}$ are assumed to be universal so that, once they are determined by data, they can be used to predict the cross-sections in other processes and in other kinematical regions. The most basic prediction of the CEM is that the ratio of the cross-sections for any two quarkonium states should be constant, independent of the process and the kinematical region. Variations in these ratios have been observed: for example the ratio of the cross-sections for $\chi_{c}$ and $J/\psi$ are rather different in photoproduction and hadroproduction and the ratio between different charmonium cross-sections measured at LHC is not constant as a function of $p_T$. These variations represent a serious challange to the status of the CEM as a quantitative phenomenological model for quarkonium production. However, the model is still widely used as simulation benchmark since, once the $F_{H}$ fractions are determined, it has a full predicting power about cross-sections but it fails to predict the quarkonium polarization. 

On the other hand, NRQCD can predict both the cross-section and the polarization of quarkonium production. In NRQCD, contributions to the quarkonium cross-section from the heavy-quark pairs produced in a color-octet (CO) state are also taken into account, in addition to the CS contributions described above.     
The picture of the NRQCD~\cite{prd51} formalism is as follows. The orbital splittings in case of quarkonium bound states are smaller than the heavy quark mass $m_{Q}$, which suggests that all the other dynamical scales of these systems are smaller than $m_{Q}$. So, the relative velocity $v$ between $Q$ and $\bar Q$ is believed to be a small quantity, $v$ $<<$ 1. Therefore, a hierarchy of scales, $m_{Q}$ $>>$ $m_{Q}v$ $>>$ $m_{Q}v^{2}$, as observed in a non-relativistic (NR) system, also holds for quarkonia. Here, $m_{Q}$ fixes the distance range for $Q\bar Q$ creation and 
annihilation processes, the momentum $m_{Q}v$ is inversely proportional to the spatial size of the bound state and the kinetic energy $m_{Q}v^{2}$ determines the typical interaction time scale. These different distance scales make the study of quarkonium production interesting and NRQCD calculation incorporates this scale hierarchy. 

The quarkonia production in NRQCD is calculated in two steps. At first, the creation of the $Q\bar Q$ pair in a hard scattering at short distances which is calculated perturbatively as an expansions in the the strong coupling constant $\alpha_{s}$. Note that $Q\bar Q$ states can be in a CS state~\cite{zpc19,npb291,plb184} as well as in a CO state~\cite{prl74,prd53,prd53a}. Then, the $Q\bar Q$ pair is evolved into the quarkonium state with the probabilities that are given by the assumed universal nonperturbative long-distance matrix elements (LDMEs) which are estimated on the basis of the comparison with experimental measurements. For CO states, this evolution process also involves the nonperturbative emission of soft gluons to form CS states. The crucial feature of this formalism is that it takes into account the complete structure of the $Q\bar Q$ Fock space, which is spanned by the states $n=\,^{2S+1}L^{[i]}_J$, where $S$, $L$ and $J$ are the spin, orbital and total angular momenta, respectively and $i$ is the color multiplicity. A remarkable progress has been made in quarkonium production studies during last decade based on the NRQCD formalism~\cite{prd78,prl100,plb673,prd84,prl106,prl106a,prl110,prl113}.

In recent times, the charmonium production in $p-p$ collisions has been measured at $\sqrt{s} = 2.76$ and 7 TeV by the ALICE~\cite{plb718,epjc}, ATLAS~\cite{npb850}, CMS~\cite{jhep02} and LHCb~\cite{epjc71,epjc72,jhep041} Collaborations at forward, near forward and mid rapidities. 
It may be noted here that ATLAS, CMS and LHCb Collaborations report the prompt production cross-sections while ALICE measurements include also the $B$ feed-down to $J/\psi$ and $\psi(2S)$. The FONLL~\cite{jhep9805,jhep0103} formalism has been used to calculate the production cross-sections of $J/\psi$ and $\psi(2S)$ from $B$ meson decays which accounts for the feed-down contributions from $B$ meson to the $J/\psi$ and $\psi(2S)$ productions.

In the present work, the charmonium cross-sections have been calculated at $\sqrt{s} = 2.76$ and 7 TeV within the framework of LO NRQCD and compared with available experimental data from LHC. The predictions for the production cross-sections of $J/\psi$ and $\psi(2S)$ in $p-p$ collisions at $\sqrt{s}$ = 13 TeV has been made as these collisions are foreseen at LHC in 2015. In addition, the calculations have also been performed at $\sqrt{s}$ = 5.1 TeV which can be utilized for the normalization of the Pb-Pb data to be collected at $\sqrt{s_{NN}} = 5.1$ TeV.

The organization of the paper is as follows. In Sec. II, we give a brief description of the theoretical model of NRQCD. Results and comparison with experimental measurements will be presented in Sec. III followed by summary and discussion in Sec. IV.

\section{$p_T$ spectrum in $\bf p\,+\,p$ collisions}
The factorization formalism of the NRQCD  provides a theoretical framework for studying the heavy quarkonium production and decay. According to the NRQCD
factorization  formalism, the cross-section for direct production of a resonance $H$ in a collision of particle $A$ and $B$ can be expressed as 
\bea
d\sigma_{A+B\rightarrow H+X} = \sum_{a,b,n}\int dx_a dx_b  G_{a/A}(x_a,\,\mu^{2}_{F})
\, G_{b/B}(x_b,\,\mu^{2}_{F})\nn\\
~~~~~~~~~~~~~~~~~~~~~\times d\sigma(a+b\rightarrow Q\bar Q(n) +X)<\mathcal{O}^H(n)>
\eea
where, $G_{a/A}(G_{b/B})$ is the parton distribution function (PDF) of the incoming parton $a(b)$ in the incident hadron $A(B)$, which depends on 
the momentum fraction $x_a(x_b)$ and the factorization scale $\mu_F$ as well as on the renormalization scale $\mu_R$. However, as we have chosen $\mu_F$ = $\mu_R$, in our case PDFs are function of $x$ and $\mu_F$ only. The tranverse mass of the resonance $H$ is $m_T = \sqrt{p_T^2 + m_H^2}$,
where $m_H\sim2m_Q$ is the mass of resonance $H$. The short distance contribution $d\sigma(a+b\rightarrow Q\bar Q(n) +X)$ 
can be calculated within the framework of perturbative QCD (pQCD). On the other hand, $<\mathcal{O}^H(n)>$ (the state $n=^{2S+1}L^{[i]}_J$) are nonperturbative LDMEs and can be estimated on the basis of the comparison with experimental measurements.

The differential cross-section for the short distance contribution i.e. the heavy quark pair production from the reaction of the type $a\,+\,b\,\rightarrow\,c\,+\,d$, where $a,\, b$ refer  to light incident partons, $c$ refers to $Q\bar Q$ pair and $d$ is the light final state parton, can be written as~\cite{rmp59}
\bea
\frac{{d\sigma}^{ab\rightarrow cd}}{dp_T\,dy} = \int dx_a\, G_{a/A}(x_a,\,\mu^{2}_{F})\, G_{b/B}(x_b,\,\mu^{2}_{F})\nn\\
~~~~~~~~~~~~\times 2p_T \frac{x_a\,x_b}{x_a-\frac{m_T}{\sqrt{s}}e^y}\frac{d\sigma}{d\hat t}(ab\rightarrow cd),
\eea
where, $\sqrt{s}$ being the total energy in the centre-of-mass and $y$ is the rapidity of the $Q\bar Q$ pair. In our numerical computation, we use CTEQ6M~\cite{prd82} for the parton distribution functions. The invariant differential cross-section is given by
\be
\frac{d\sigma}{d\hat t} = \frac{|\mathcal{M}|^2}{16\pi{\hat s}^2},
\ee
where $\hat s$ and $\hat t$ are the parton level Mandelstam variables. $\mathcal{M}$ is the feynman amplitude for the process. The value of the momentum fraction $x_b$ can be written as,
\be
x_b = \frac{1}{\sqrt{s}}\frac{x_a\,\sqrt{s}\,m_T\,e^{-y}-m^2_H}{x_a\,\sqrt{s}-m_T\,e^y}.
\ee
The minimum value of $x_a$ is 
\be
x_{a\rm min} = \frac{1}{\sqrt{s}}\frac{\sqrt{s}\,m_T\,e^{y}-m^2_H}{\sqrt{s}-m_T\,e^{-y}}.
\ee

The LDMEs are predicted to scale with a definite power of the relative velocity $v$ of the heavy constituents inside $Q\bar Q$ bound states. In the limit $v<<1$, the production of quarkonium is based on the $^3S_1^{[1]}$ and $^3P_J^{[1]}$ ($J$ = 0,1,2) CS states and $^1S_0^{[8]}$, $^3S_1^{[8]}$ and $^3P_J^{[8]}$ CO states. In our calculations, we used the expressions for the short distance CS cross-sections given in Refs.~\cite{zpc19,npb291,plb184} and the CO cross-sections given in Refs.~\cite{prd53,prd53a}.

In this paper we calculate the $p_T$ distribution of $J/\psi$ and $\psi(2S)$ in $p-p$ collisions at LHC energies. For $J/\psi$ production in $p-p$ collisions, three sources need to be considered: direct $J/\psi$ production, feed-down contributions to the $J/\psi$ from the decay of heavier charmonium states, predominantly from $\psi(2S)$, $\chi_{c0}$, $\chi_{c1}$ and $\chi_{c2}$ and $J/\psi$ from $B$ hadron decays. The sum of the first two sources is called "prompt $J/\psi$" and the third source will be called "$J/\psi$ from $B$". On the other hand, $\psi(2S)$ has no significant feed-down contributions from higher mass states. We call this direct contribution as "prompt $\psi(2S)$" to be consistent with the experiments. The other source to $\psi(2S)$ production is from $B$ hadron decays and we call it "$\psi(2S)$ from $B$". The sum of the prompt $J/\psi$($\psi(2S)$) and $J/\psi$($\psi(2S)$) from $B$ will be called "inclusive $J/\psi$($\psi(2S)$)".


The direct production cross-section of $J/\psi$ can be written as the sum of the contributions~\cite{prd53,prd53a}, 
\bea
d\sigma(J/\psi) = d\sigma(Q\bar Q(~^3S_1^{[1]}))<\mathcal{O}(Q\bar Q(~^3S_1^{[1]})\rightarrow J/\psi)>\nn\\
~~~~~~~~~~~~~+d\sigma(Q\bar Q(~^1S_0^{[8]}))<\mathcal{O}(Q\bar Q(~^1S_0^{[8]})\rightarrow J/\psi)>\nn\\
~~~~~~~~~~~~~+d\sigma(Q\bar Q(~^3S_1^{[8]}))<\mathcal{O}(Q\bar Q(~^3S_1^{[8]})\rightarrow J/\psi)>\nn\\
~~~~~~~~~~~~~+d\sigma(Q\bar Q(~^3P_J^{[8]}))<\mathcal{O}(Q\bar Q(~^3P_J^{[8]})\rightarrow J/\psi)>\nn\\
~~~~~~~~~~~~~+...
\eea
Similar expression holds for direct $\psi(2S)$ production. The direct production cross-section for $\chi_{cJ}$ can be written as~\cite{prd53}:
\bea
d\sigma(\chi_{cJ}) = d\sigma(Q\bar Q(~^3P_J^{[1]}))<\mathcal{O}(Q\bar Q(~^3P_J^{[1]})\rightarrow \chi_{cJ})>\nn\\
~~~~~~~~~~~~~+d\sigma(Q\bar Q(~^3S_1^{[8]}))<\mathcal{O}(Q\bar Q(~^3S_1^{[8]})\rightarrow \chi_{cJ}>\nn\\
~~~~~~~~~~~~~+....
\eea
Here, we have taken into account the contributions from all three $\chi_{cJ}$ ($\chi_{c0}$, $\chi_{c1}$ and $\chi_{c2}$) mesons to $J/\psi$.

\begin{table}[t]
\centering
\begin{tabular}{l|cccc}
\hline
\hline
~&~~~~~~~~~~~~~~~~~~~~~~~&~~~~~~~~~~~~~~~~~~~~&NRQCD\\
&LDMEs&Numerical&scaling\\
&~~~&value&order\\
\hline
&~$<\mathcal{O}(Q\bar Q(~^3S_1^{[1]})\rightarrow J/\psi)>$ & 1.2 GeV$^{3}$& $m_{c}^{3}v_{c}^{3}$ \\  
Color-&~$<\mathcal{O}(Q\bar Q(~^3S_1^{[1]})\rightarrow \psi(2S))>$ & 0.76 GeV$^{3}$& $m_{c}^{3}v_{c}^{3}$ \\ 
Singlet&~$<\mathcal{O}(Q\bar Q(~^3P_{0}^{[1]}) \rightarrow \chi_{c0})>$/$m_{c}^{2}$ & 0.054 GeV$^{3}$& $m_{c}^{3}v_{c}^{5}$ \\ 
&~$<\mathcal{O}(Q\bar Q(~^3P_{1}^{[1]}) \rightarrow \chi_{c1})>$/3$m_{c}^{2}$ & 0.054 GeV$^{3}$& $m_{c}^{3}v_{c}^{5}$ \\
&~$<\mathcal{O}(Q\bar Q(~^3P_{2}^{[1]}) \rightarrow \chi_{c2})>$/5$m_{c}^{2}$ & 0.054 GeV$^{3}$& $m_{c}^{3}v_{c}^{5}$ \\
\hline
&~$<\mathcal{O}(Q\bar Q(~^3S_1^{[8]})\rightarrow J/\psi)>$ & 0.0013 $\pm$ 0.0013 GeV$^{3}$& $m_{c}^{3}v_{c}^{7}$ \\
&~$<\mathcal{O}(Q\bar Q(~^3S_1^{[8]})\rightarrow \psi(2S))>$ & 0.0033 $\pm$ 0.00021 GeV$^{3}$& $m_{c}^{3}v_{c}^{7}$ \\
Color-&~$<\mathcal{O}(Q\bar Q(~^3S_1^{[8]})\rightarrow \chi_{cJ})>$/$m_{c}^{2}$ & 0.00187 $\pm$ 0.00025 GeV$^{3}$& $m_{c}^{3}v_{c}^{5}$ \\
Octet&~$<\mathcal{O}(Q\bar Q(~^1S_0^{[8]})\rightarrow J/\psi)>$ & 0.018 $\pm$ 0.0087 GeV$^{3}$& $m_{c}^{3}v_{c}^{7}$ \\
&~$<\mathcal{O}(Q\bar Q(~^1S_0^{[8]})\rightarrow \psi(2S))>$ & 0.0080 $\pm$ 0.00067 GeV$^{3}$& $m_{c}^{3}v_{c}^{7}$ \\
&~$<\mathcal{O}(Q\bar Q(~^3P_0^{[8]})\rightarrow J/\psi)>$/$m_{c}^{2}$ & 0.018 $\pm$ 0.0087 GeV$^{3}$& $m_{c}^{3}v_{c}^{7}$ \\
&~$<\mathcal{O}(Q\bar Q(~^3P_0^{[8]})\rightarrow \psi(2S))>$/$m_{c}^{2}$ & 0.0080 $\pm$ 0.00067 GeV$^{3}$& $m_{c}^{3}v_{c}^{7}$ \\
\hline
\hline
\end{tabular}
\caption{The color-singlet and color-octet matrix elements with numerical values and NRQCD scaling order.}
\label{tab:h}
\end{table}

To calculate the direct charmonia and feed-down contributions from heavier states as well as from $B$ decays, we use the following branching ratios:
$\mathcal{B}$($J/\psi[\psi(2S)]\rightarrow \mu^{+}\mu^{-}$)=0.0593[0.0078], 
$\mathcal{B}$[$\psi(2S)\rightarrow J/\psi$]=0.603. 
$\mathcal{B}$($\chi_{cJ}\rightarrow J/\psi$)=0.0130, 0.348, 0.198 for $J$ = 0, 1, 2, respectively and
$\mathcal{B}$($B\rightarrow J/\psi[\psi(2S)]$)=0.116[0.283]~\cite{prd86}. To choose the renormalization scale $\mu_{R}$ and the factorization scale $\mu_F$ in this calculations is an important issue and it may cause the uncertainties in the calculations. The choice that $\mu_{F}$ = $\mu_{R}$ = $\sqrt{p_T^2 + 4m_c^2}$ is the default one in the calculation, with $m_c$ being mass of the charm quark assumed to be 1.4 GeV. Moreover, it has been shown in~\cite{prd78} that the scale variation does not improve the result for $J/\psi$. Thus, in our case we kept $\mu_{F}$ = $\mu_{R}$. 
The LDMEs~\cite{prc87} for CS and CO which we have used for our calculations are given in the Table I. The central values of LDMEs are taken for our calculations. For FONLL~\cite{jhep9805,jhep0103,cacci} calculations, PDF used is CTEQ6.6 and the central values of the factorization and renormalization scales are chosen to be $\mu$ = $\sqrt{p_T^2 + m_b^2}$, where $p_T$ and $m_b$ are the transverse momentum and mass of the b-quark and central value of $m_b$ = 4.75 GeV is used.

In order to estimate the uncertainty on the calculated values, four possible sources have been considered namely, the factorization scale, the mass of the charm quark, the branching ratios for the feed-down to $J/\psi$ and the PDFs. The largest uncertainty in the branching ratio is 5\% which corresponds to $\chi_{cJ}\rightarrow J/\psi + \gamma$ channel. The uncertainty due to the assumed PDF was estimated by performing the calculations with different PDFs namely, CTEQ6M, CTEQ6L and CTEQ6L1. The results of these calculations agree within 10\%. The uncertainty due to the charm quark mass was estimated by carrying out the numerical calculations for $m_c$ = 1.2 and 1.6 GeV. The variation was found to be about 12\%. On the other hand, the uncertainty due to the variation in the values of factorization and renormalization scales was found to be as large as 45\% and 30\% when the value is reduced and enhanced by a factor of two, repectively. Thus, this is the most dominant source of uncertainy and in all the subsequent plots for the numerical values, the uncertainty bands correspond only to this source. This assumption is valid in case the four sources of uncertainty are assumed to be uncorrelated and can be added in quadrature.


\section{Results}

The NRQCD calculations have been carried out for the differential cross-sections of $J/\psi$ and $\psi(2S)$ as a function of $p_T$ at $\sqrt{s}$ = 2.76 and 7 TeV. The numerical results have been compared with experimental data available from CMS ($|y|<0.9$, $0.9<|y|<2.4$, $|y|<1.2$ and $1.2<|y|<2.4$), ATLAS ($|y|<0.75$ and $0.75<|y|<2.4$), LHCb ($2<y<4.5$) and ALICE ($2.5<y<4$). Thus, this comprehensive study explores the validity of NRQCD calculation at mid, near forward and forward rapidities at LHC energies.

\begin{figure}[t]
\begin{center}
\includegraphics[width=6.15cm,height=5.0cm,angle=0]{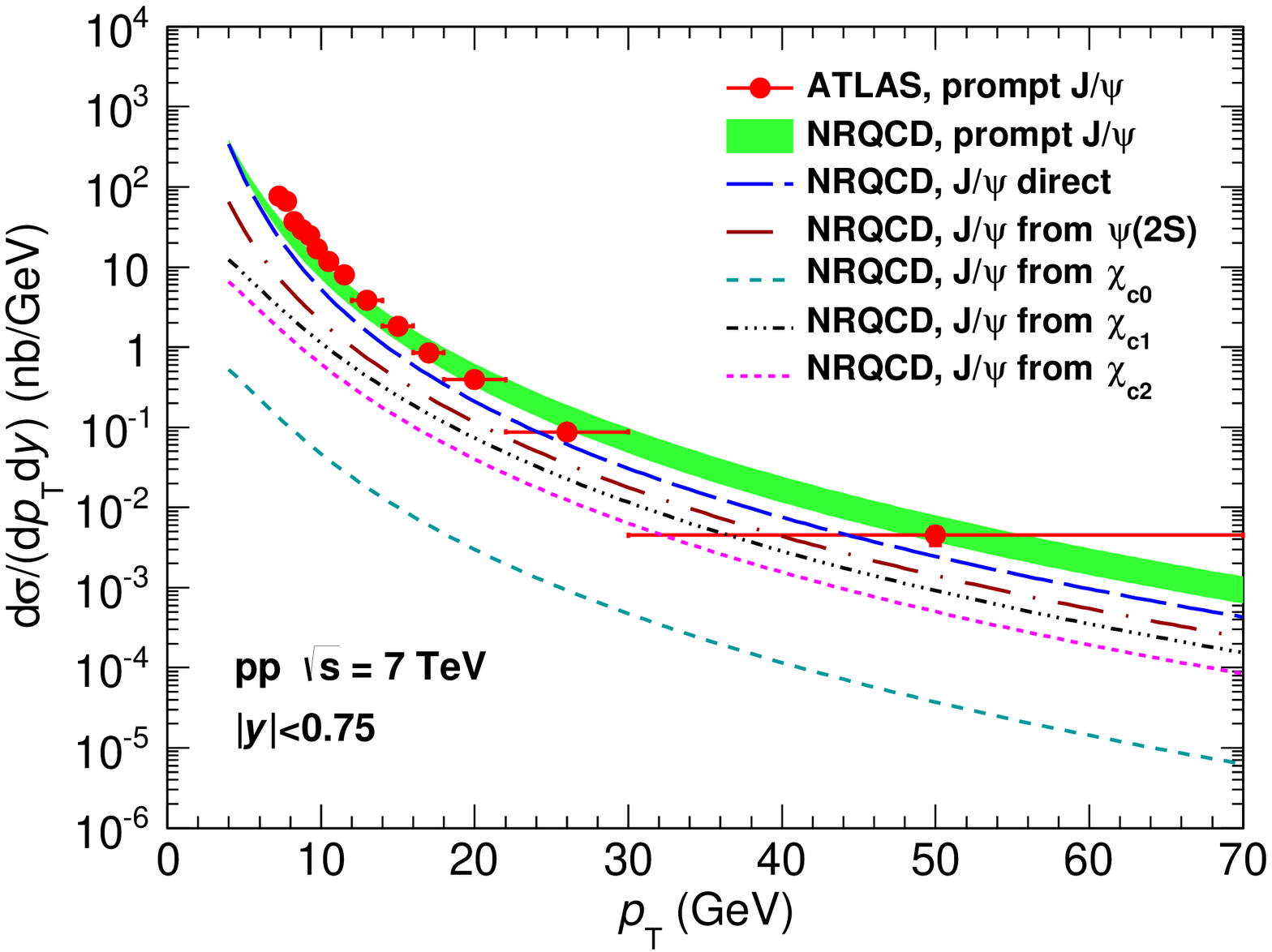}~~~~\includegraphics[width=6.15cm,height=5.0cm,angle=0]{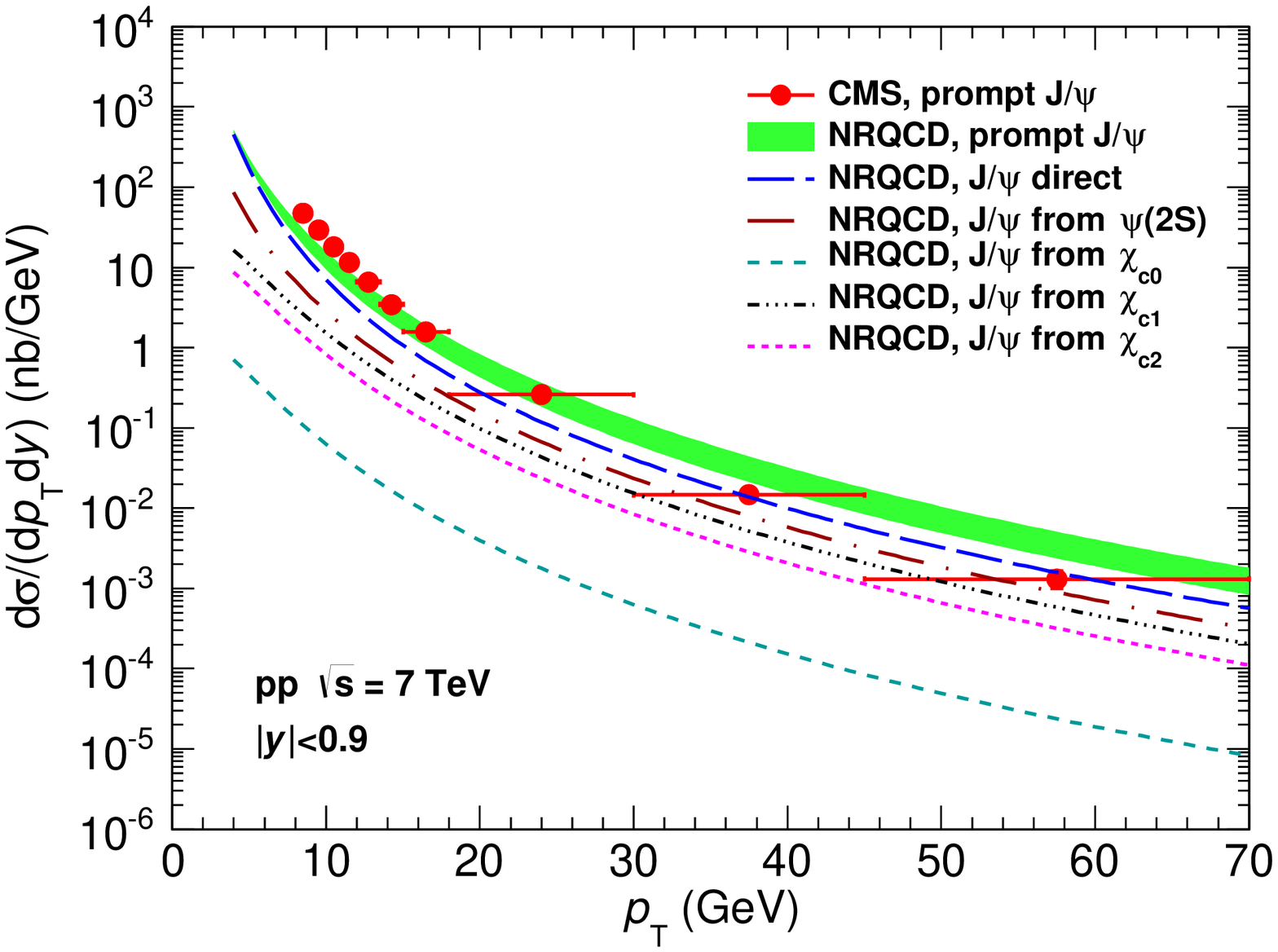}~~~~\\ \includegraphics[width=6.15cm,height=5.0cm,angle=0]{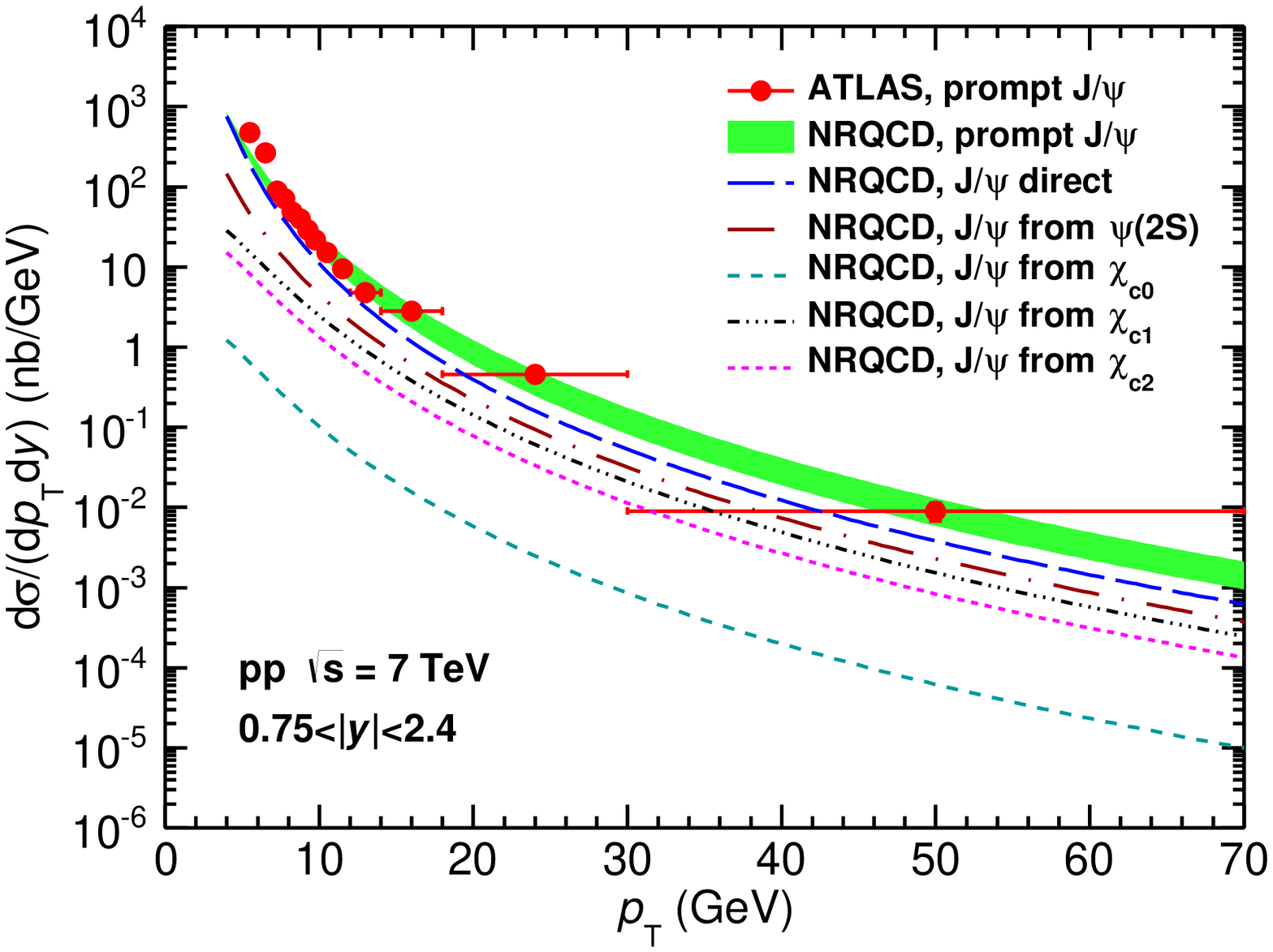}~~~~\includegraphics[width=6.15cm,height=5.0cm,angle=0]{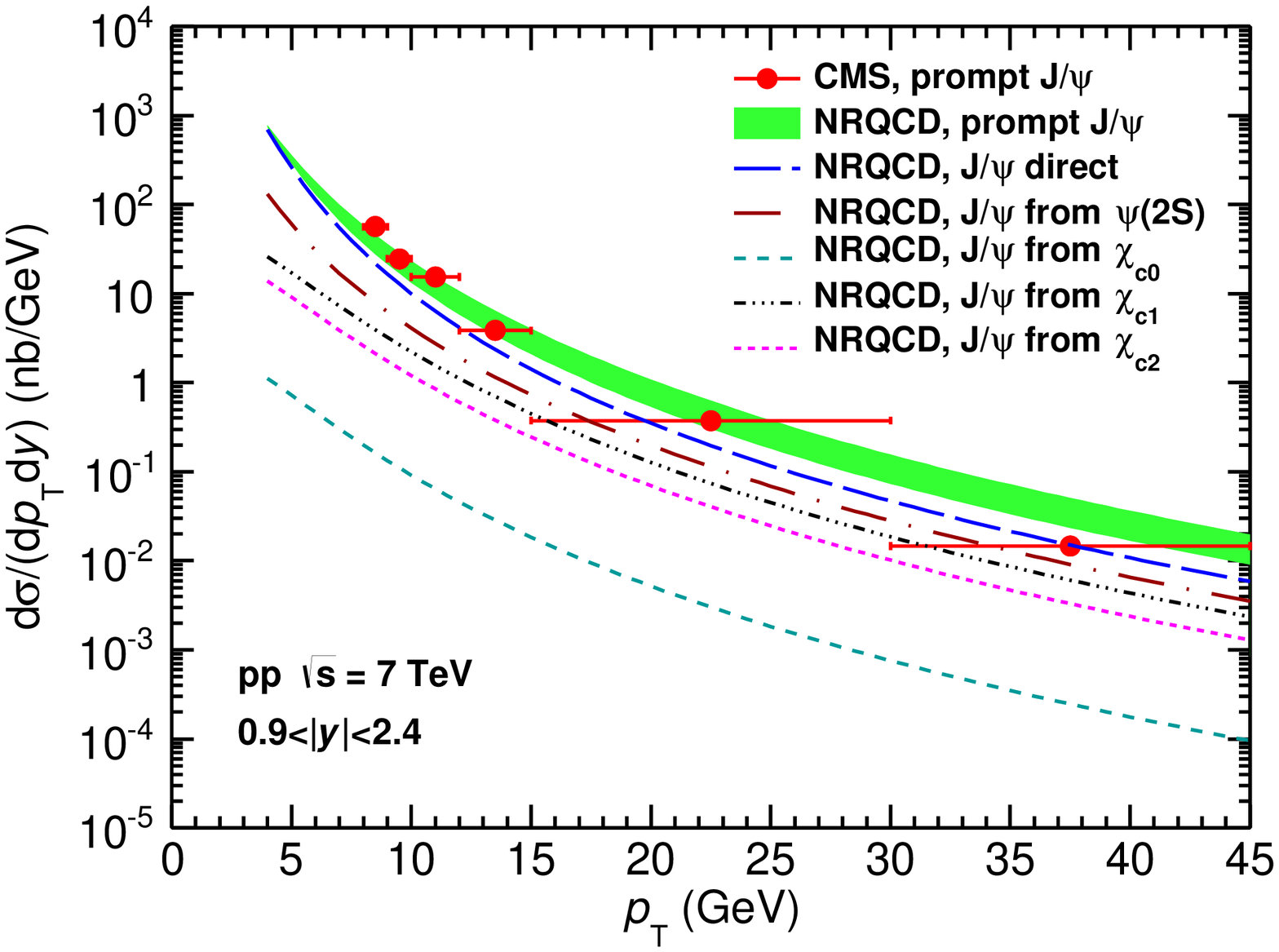}~~~~\\ \includegraphics[width=6.15cm,height=5.0cm,angle=0]{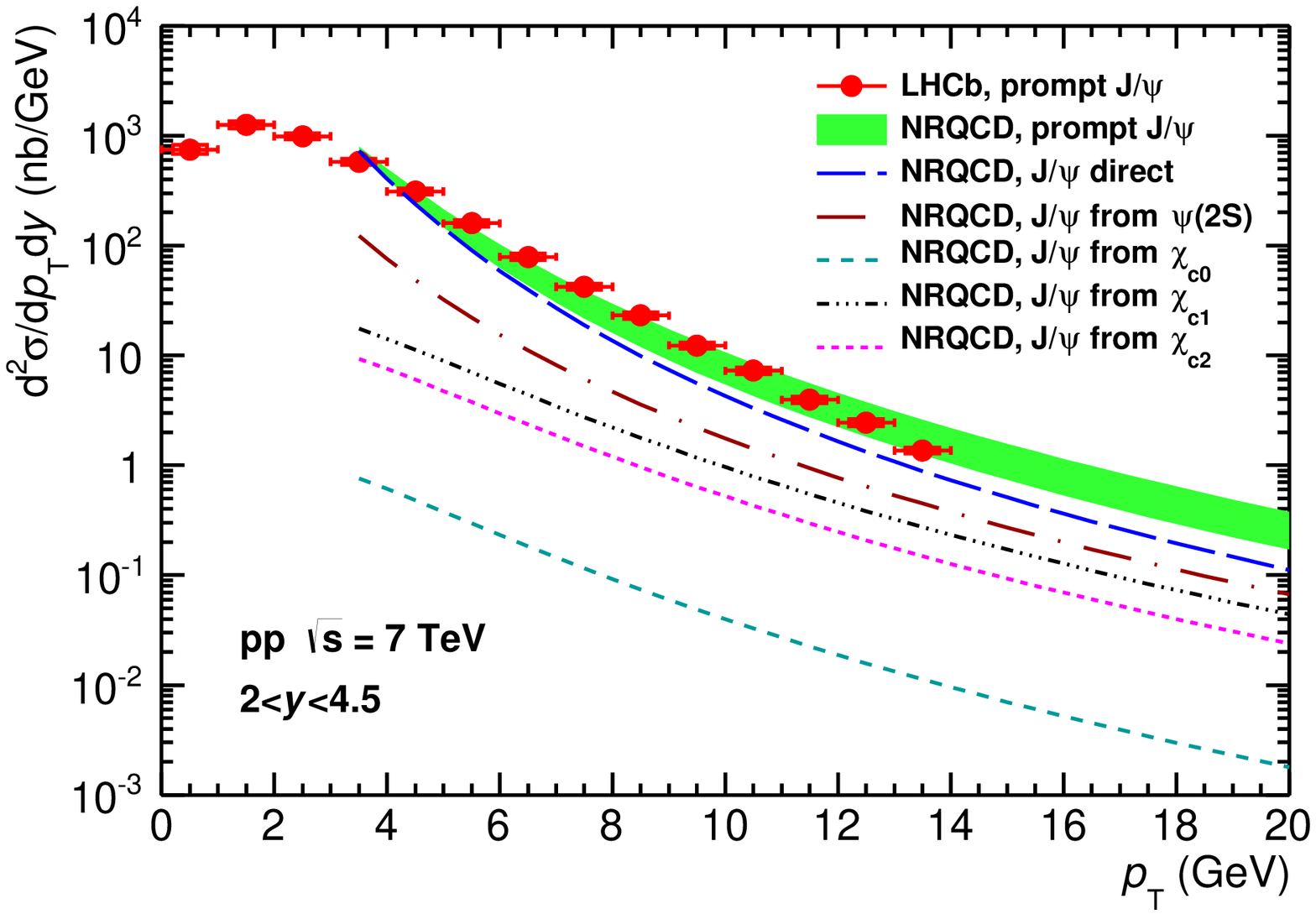}~~~~\includegraphics[width=6.15cm,height=5.0cm,angle=0]{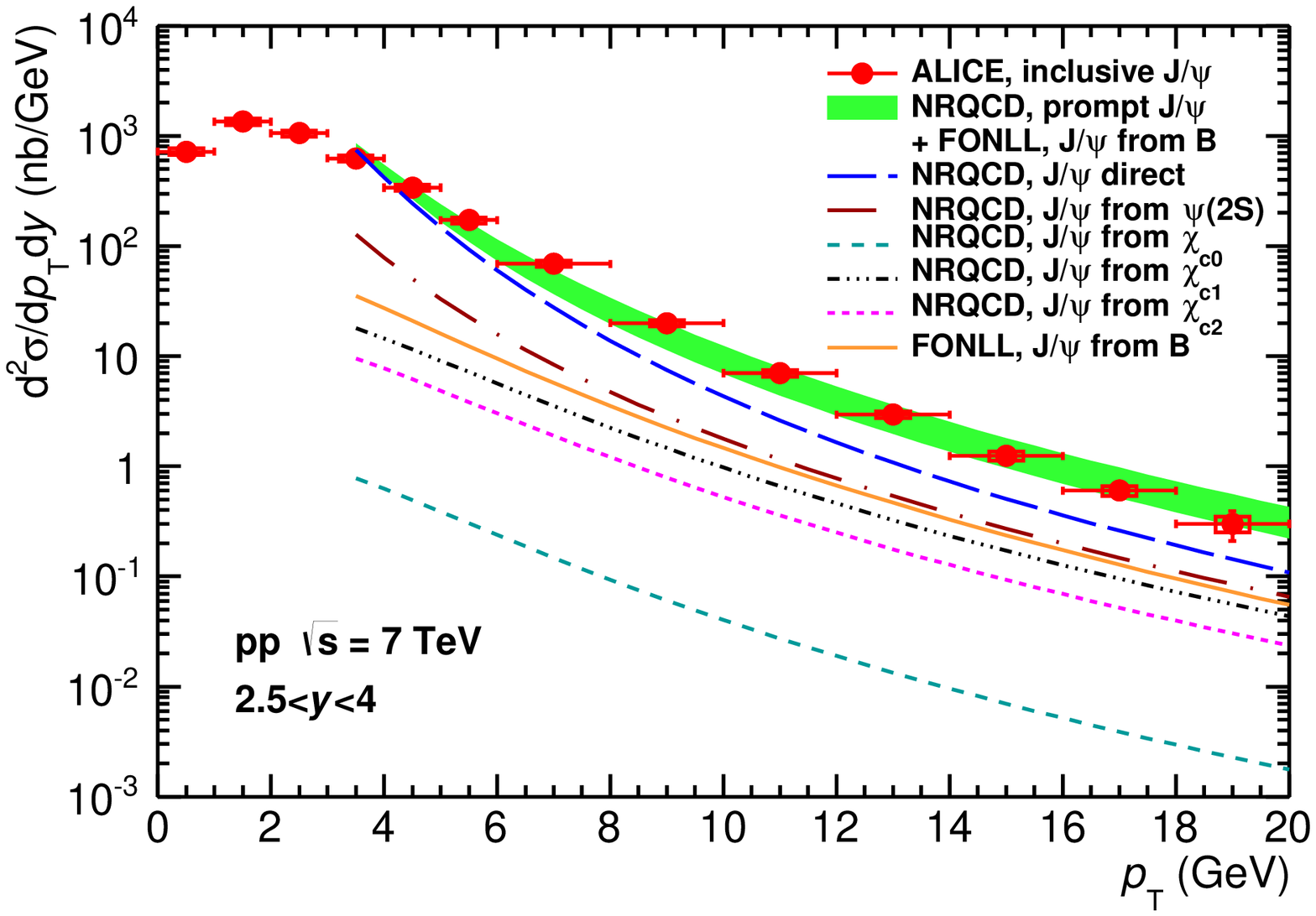}
\end{center}
\caption{(Color online) Differential production cross-section of $J/\psi$ as a function of $p_{T}$ compared with the ATLAS~\cite{npb850}, CMS~\cite{jhep02}, LHCb~\cite{epjc71} and ALICE~\cite{plb718,epjc} data. For data, the vertical error bars represent the statistical errors while the boxes correspond to the systematic uncertainties. We have shown the sum of all contributions with a green band. The direct and feed-down contributions to $J/\psi$ are shown only by lines which are for the central values.}
\label{fig1}
\end{figure}

In Fig.~\ref{fig1}, the numerical values from the NRQCD calculations for differential cross-section of $J/\psi$ as a function of $p_T$ have been compared with the experimental values obtained by the four experiments at LHC namely, ATLAS~\cite{npb850}, CMS~\cite{jhep02}, LHCb~\cite{epjc71} and ALICE~\cite{epjc} at $\sqrt{s}$ = 7 TeV. It may be noted that the $B$ feed-down contribution in case of ALICE has been accounted using FONLL. It is observed from Fig.~\ref{fig1}, that the calculated values show good agreement with all the experimental data for $p_T>$ 4 GeV. In a recent publication by Yan-Qing Ma and Raju Venugopalan~\cite{raju}, it has been demonstrated that the low $p_T$ cross-section can be reproduced by inclusion of Color Glass Condensate (CGC) effects within the NRQCD framework.

\begin{figure}[t]
\begin{center}
\includegraphics[width=6.15cm,height=5.0cm,angle=0]{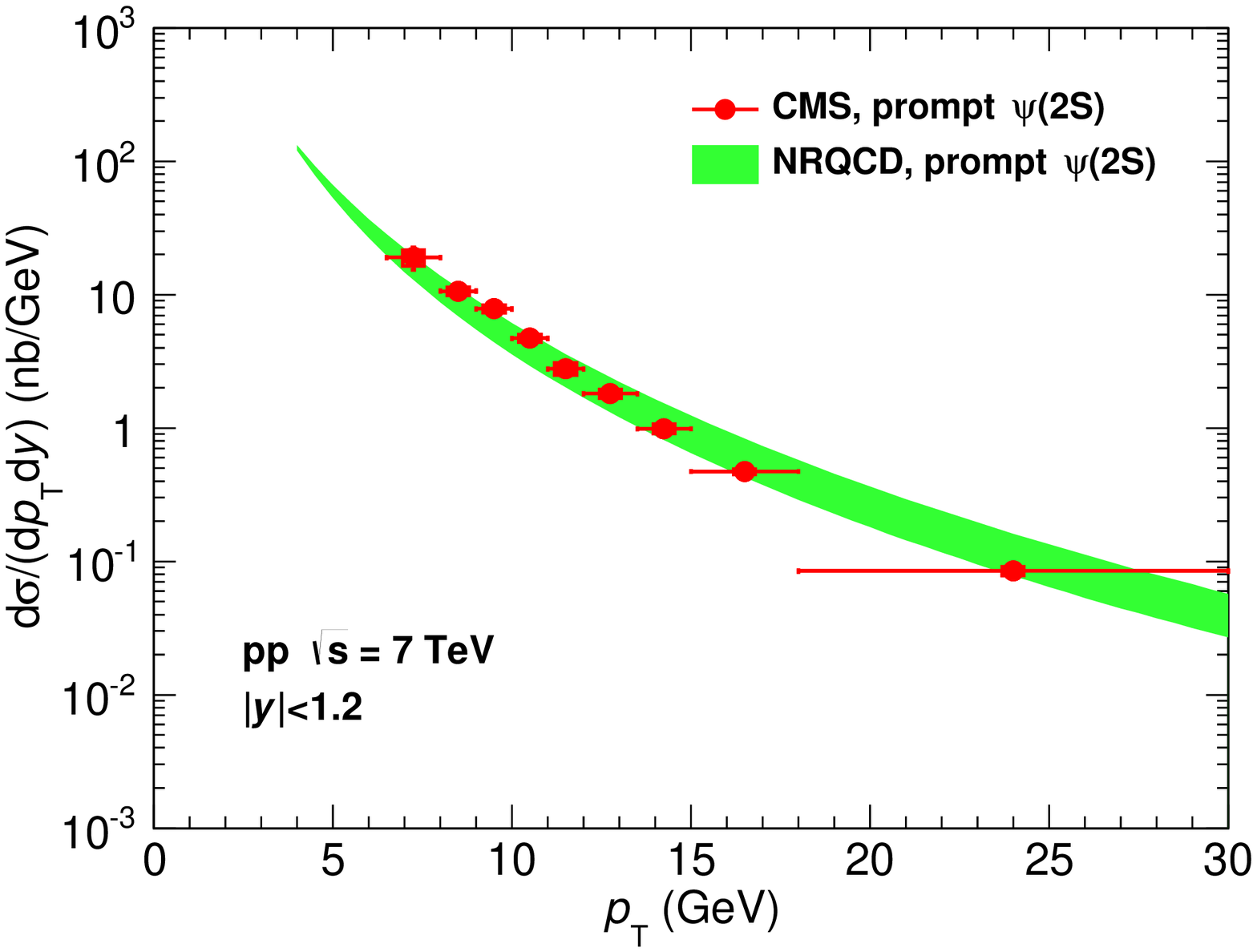}~~~~ \includegraphics[width=6.15cm,height=5.0cm,angle=0]{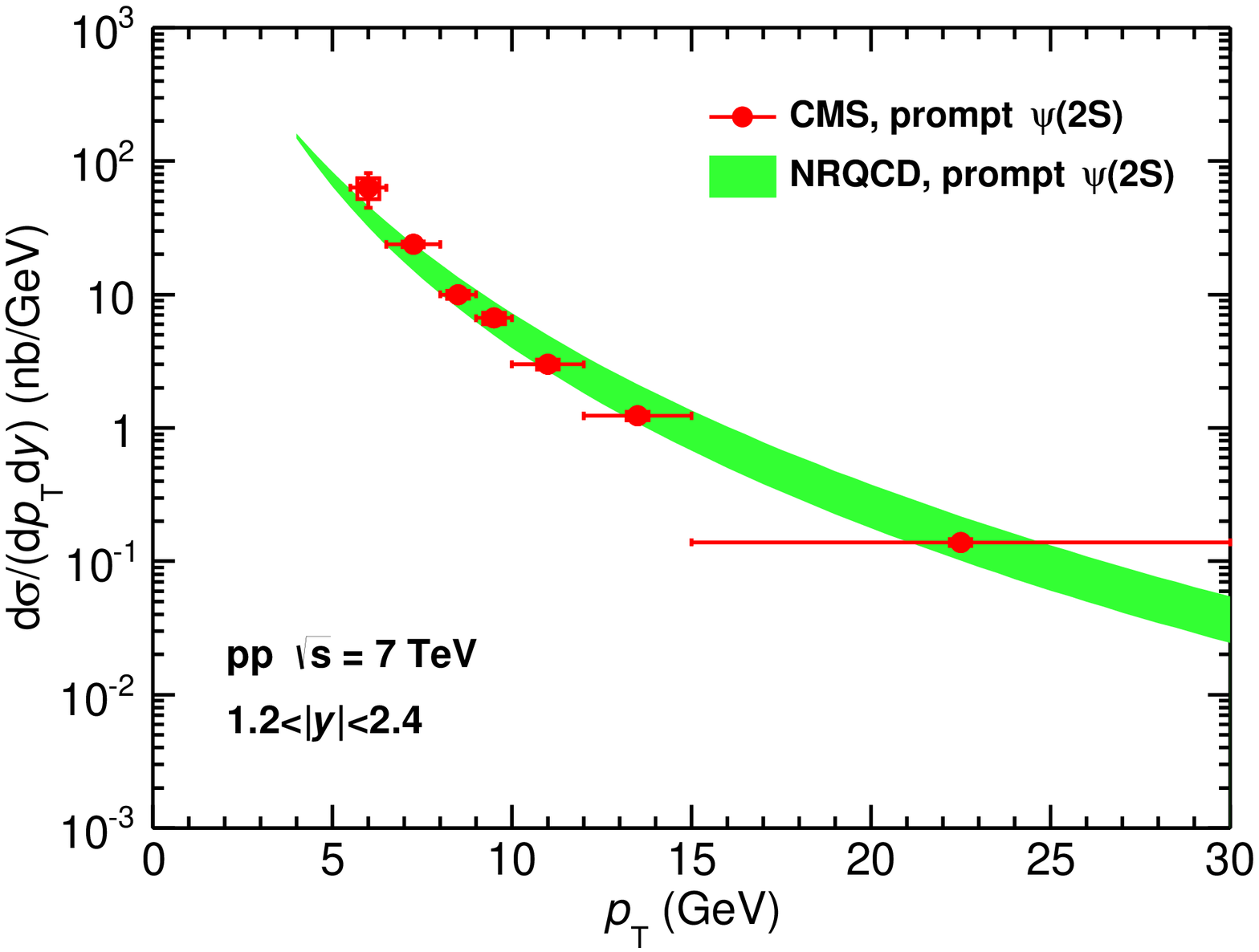}~~~~\\ \includegraphics[width=6.15cm,height=5.0cm,angle=0]{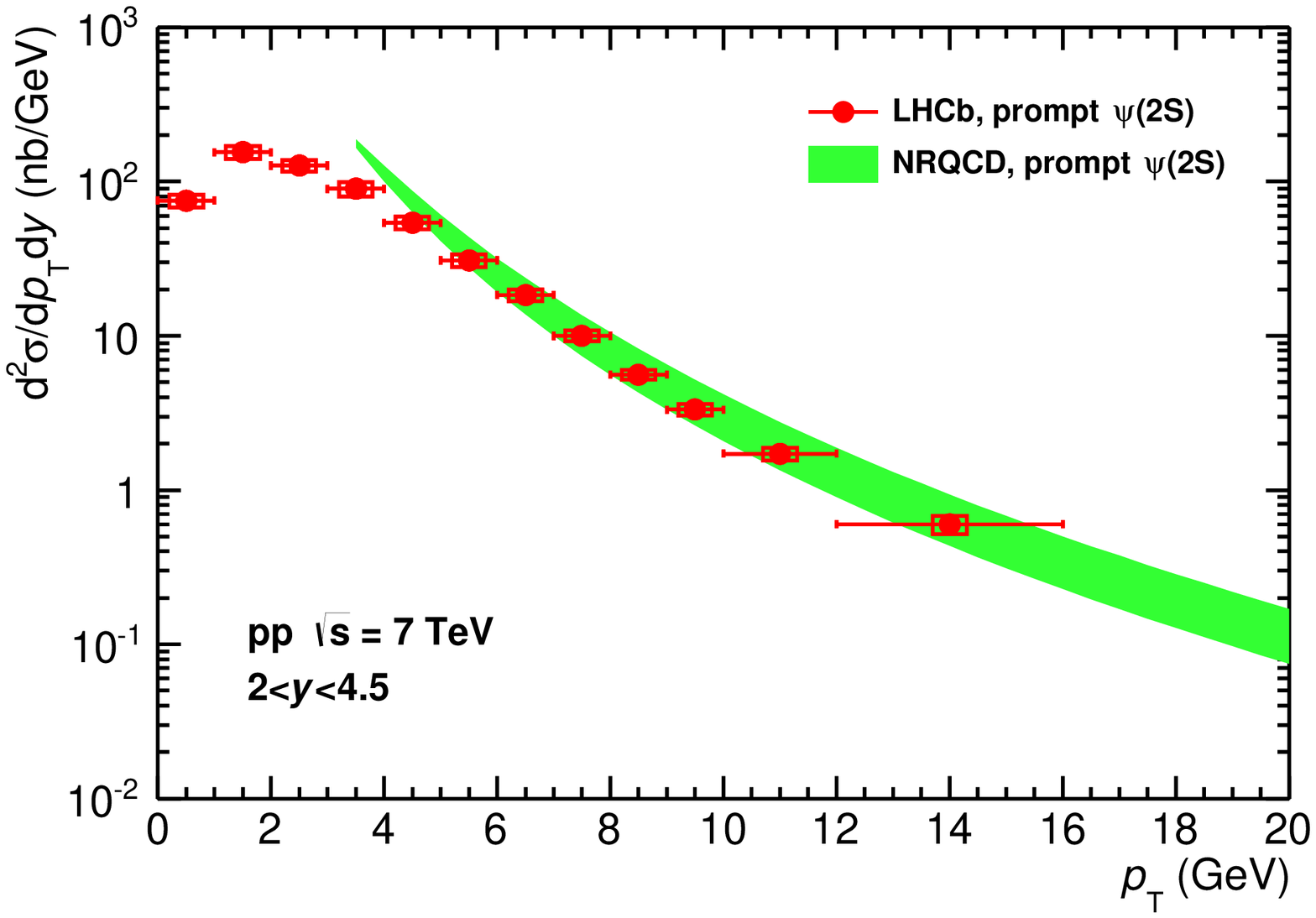}~~~~ \includegraphics[width=6.15cm,height=5.0cm,angle=0]{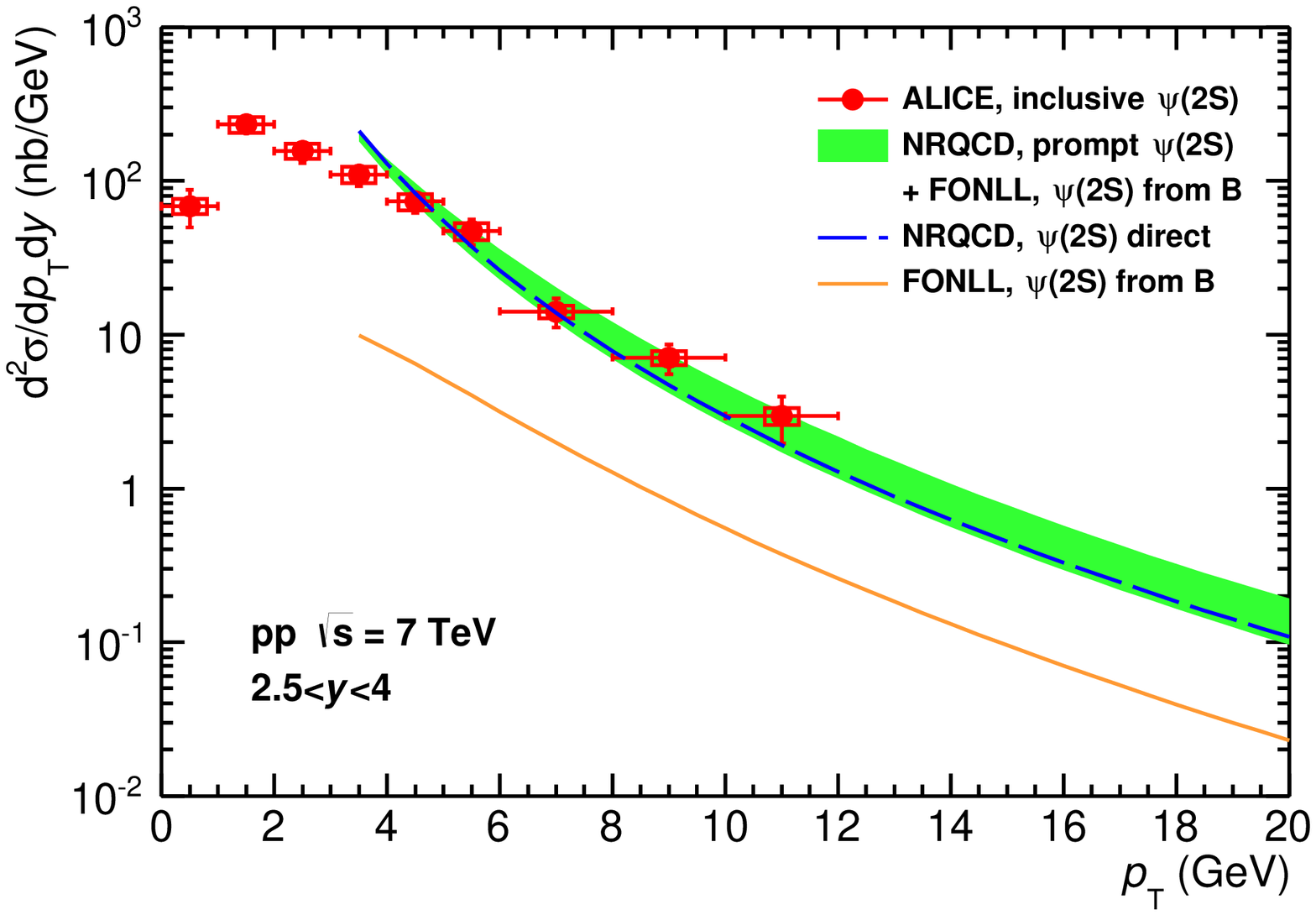}
\end{center}
\caption{(Color online) Differential production cross-section of $\psi(2S)$ as a function of $p_{T}$ compared with the 
CMS~\cite{jhep02}, LHCb~\cite{epjc72} and ALICE~\cite{epjc}.}
\label{fig2}
\end{figure}

The calculated values of the differential cross-section of $\psi(2S)$ as a function of $p_T$ using the NRQCD framework have been compared with CMS~\cite{jhep02}, LHCb~\cite{epjc72} and ALICE~\cite{epjc} and are shown in Fig.~\ref{fig2}. It is important to note that for $\psi(2S)$ there is no contribution from the higher excited charmonium states. Thus, the prompt and direct production is the same. Again for ALICE, $B$ feed-down to $\psi(2S)$ has been calculated from the FONLL. The calculated and measured values for $\psi(2S)$ are also in good agreement.

\begin{figure}[t]
\begin{center}
\includegraphics[width=6.15cm,height=5.0cm,angle=0]{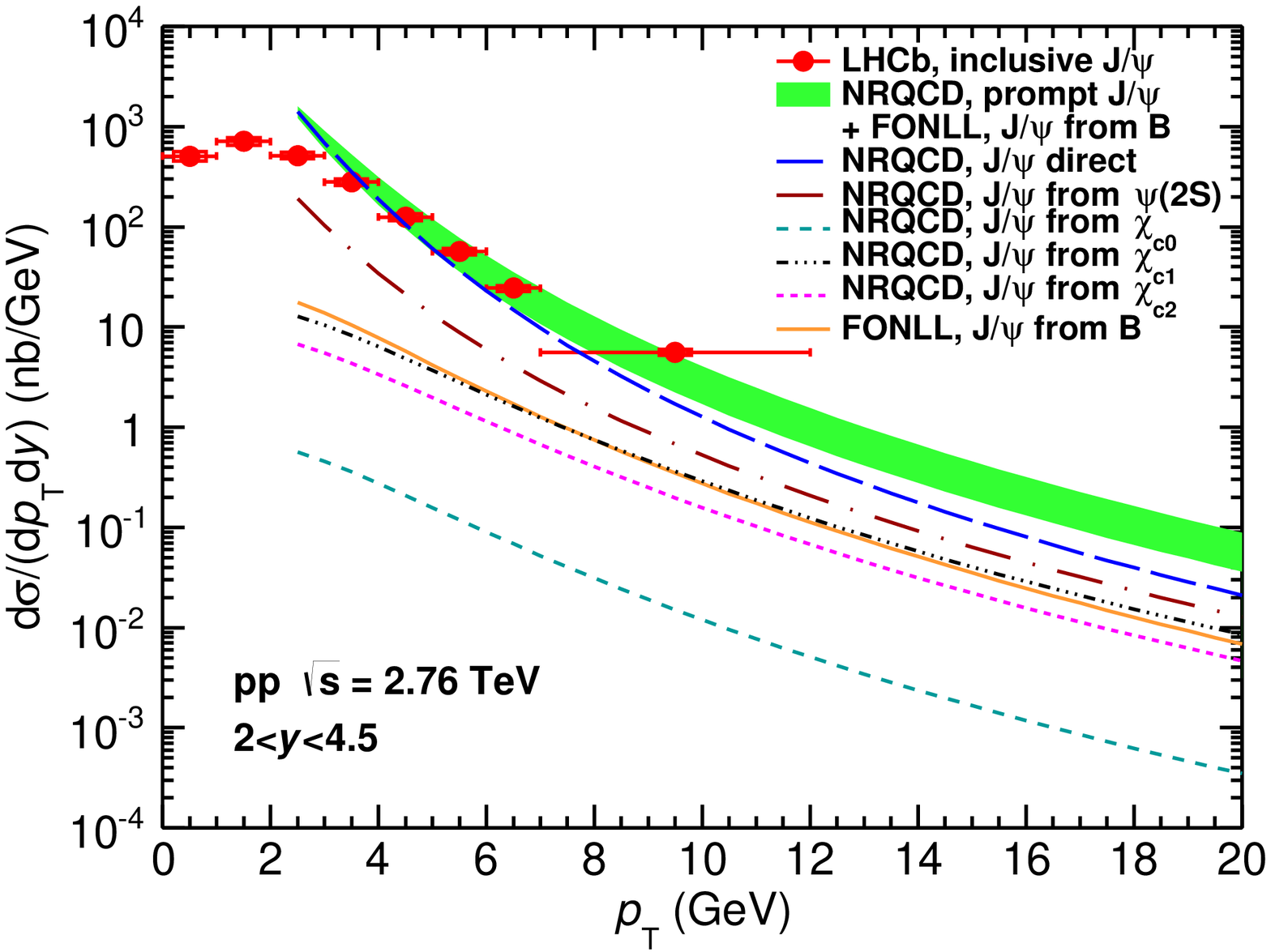}~~~~ \includegraphics[width=6.15cm,height=5.0cm,angle=0]{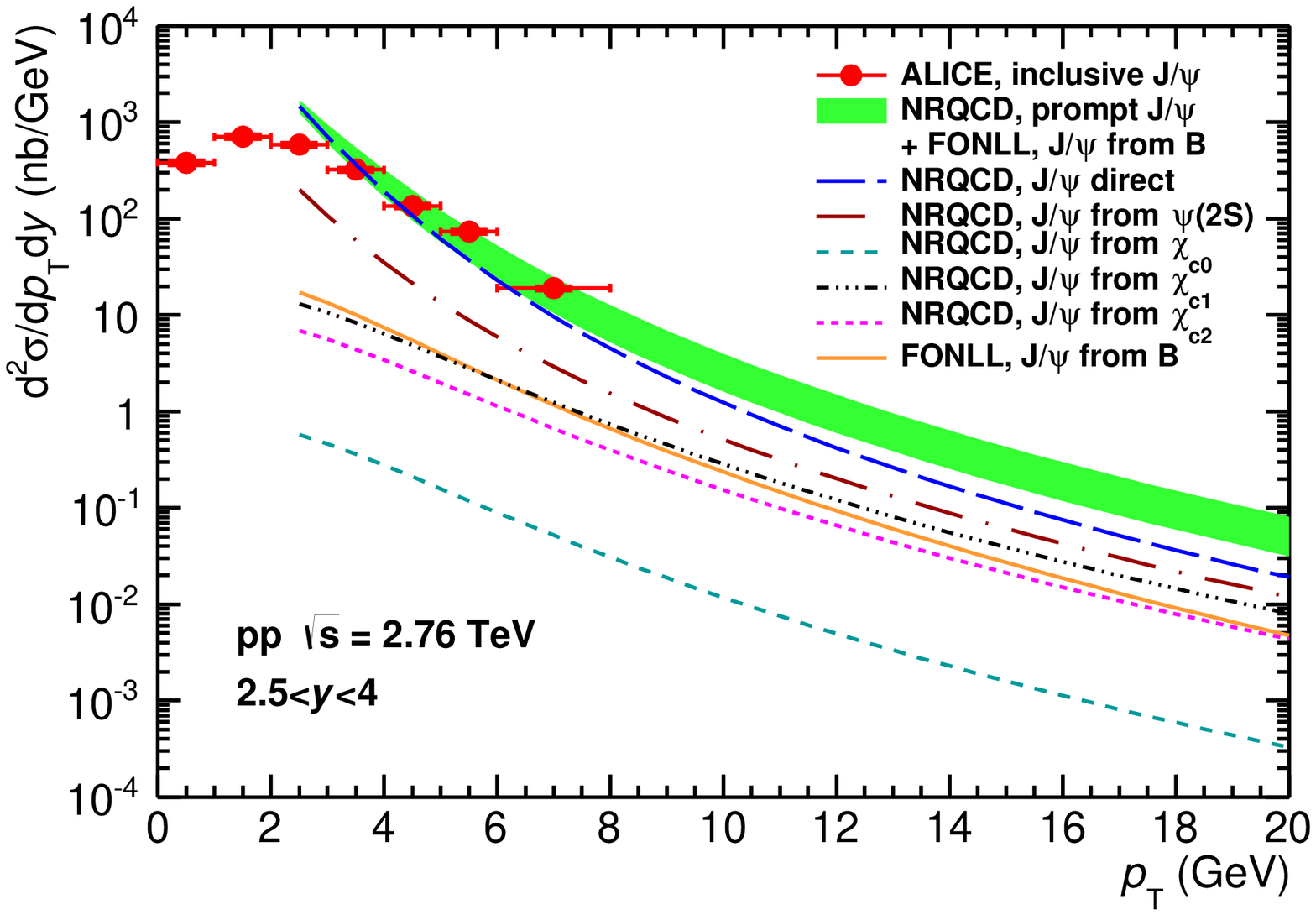}
\end{center}
\caption{(Color online) Differential production cross-section of inclusive $J/\psi$ as a function of $p_{T}$ compared with the 
LHCb~\cite{jhep041} and ALICE~\cite{epjc}.}
\label{fig3}
\end{figure}

The numerical calculations for inclusive $J/\psi$ production were also carried out for $\sqrt{s}$ = 2.76 TeV and compared with the reported inclusive measurements from LHCb~\cite{jhep041} and ALICE~\cite{plb718} in Fig.~\ref{fig3}. In this case the calculated and measured values for $J/\psi$ are in good agreement for $p_T >$ 3 GeV.

\begin{figure}[t]
 \begin{center}
 \includegraphics[width=6.5cm,height=5.0cm,angle=0]{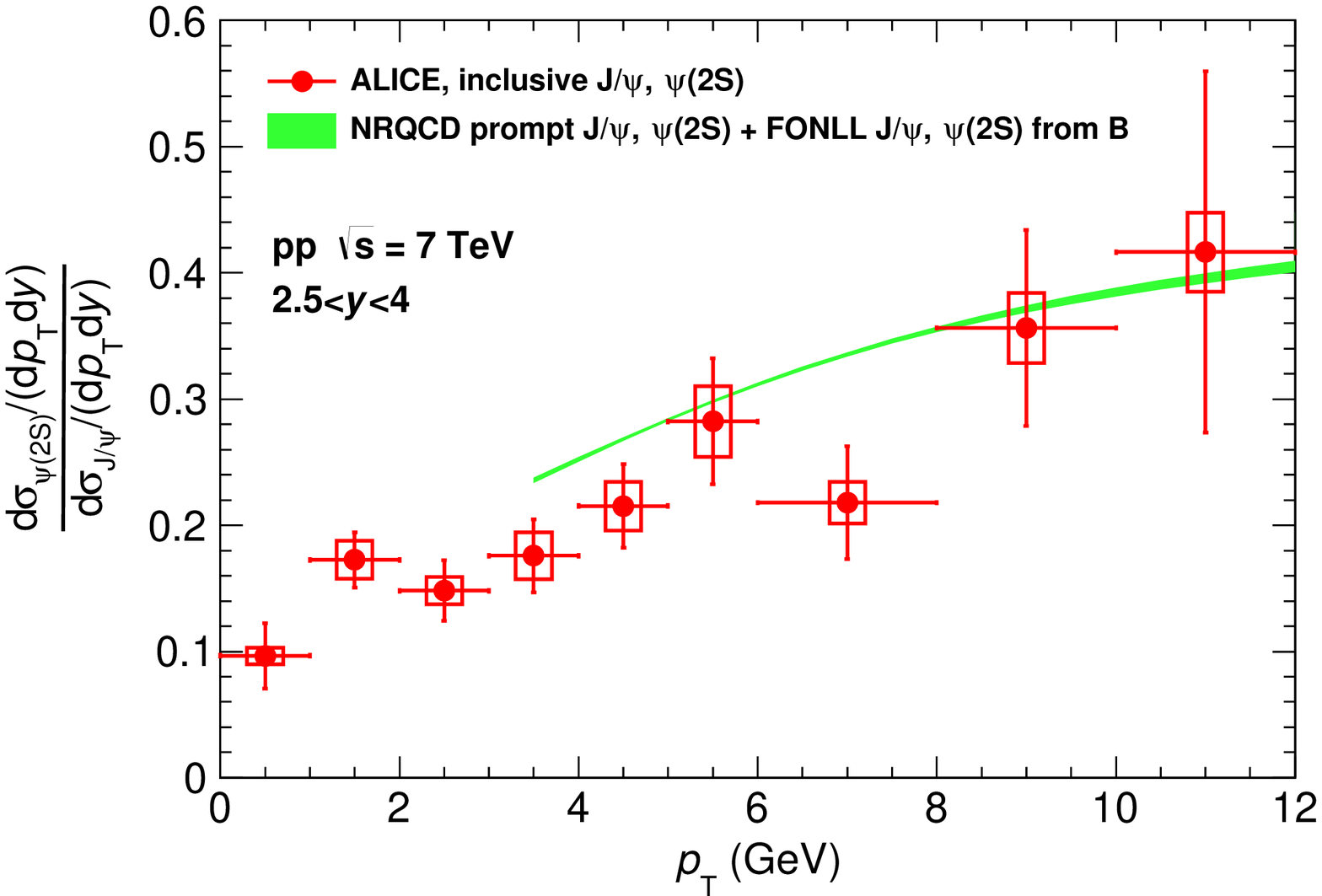}
 \end{center}
 \caption{(Color online) Inclusive $\psi(2S)$ to $J/\psi$ production cross-section ratio as a function of $p_{T}$ compared to the 
 ALICE~\cite{epjc}.}
 \label{fig4}
 \end{figure}

The ALICE Collaboration has also reported the ratio of the differential cross-sections of $\psi(2S)$ to $J/\psi$ at $\sqrt{s}$ = 7 TeV~\cite{epjc}. 
The measured and calculated values are shown in Fig.~\ref{fig4}. The agreement is reasonable for $p_{T} >$ 4 GeV and the increasing trend of the value for the ratio has been well reproduced. It is worth noting that the prediction for this ratio from CEM is independent of $p_T$. 

\begin{figure}[t]
 \begin{center}
 \includegraphics[width=6.15cm,height=5.0cm,angle=0]{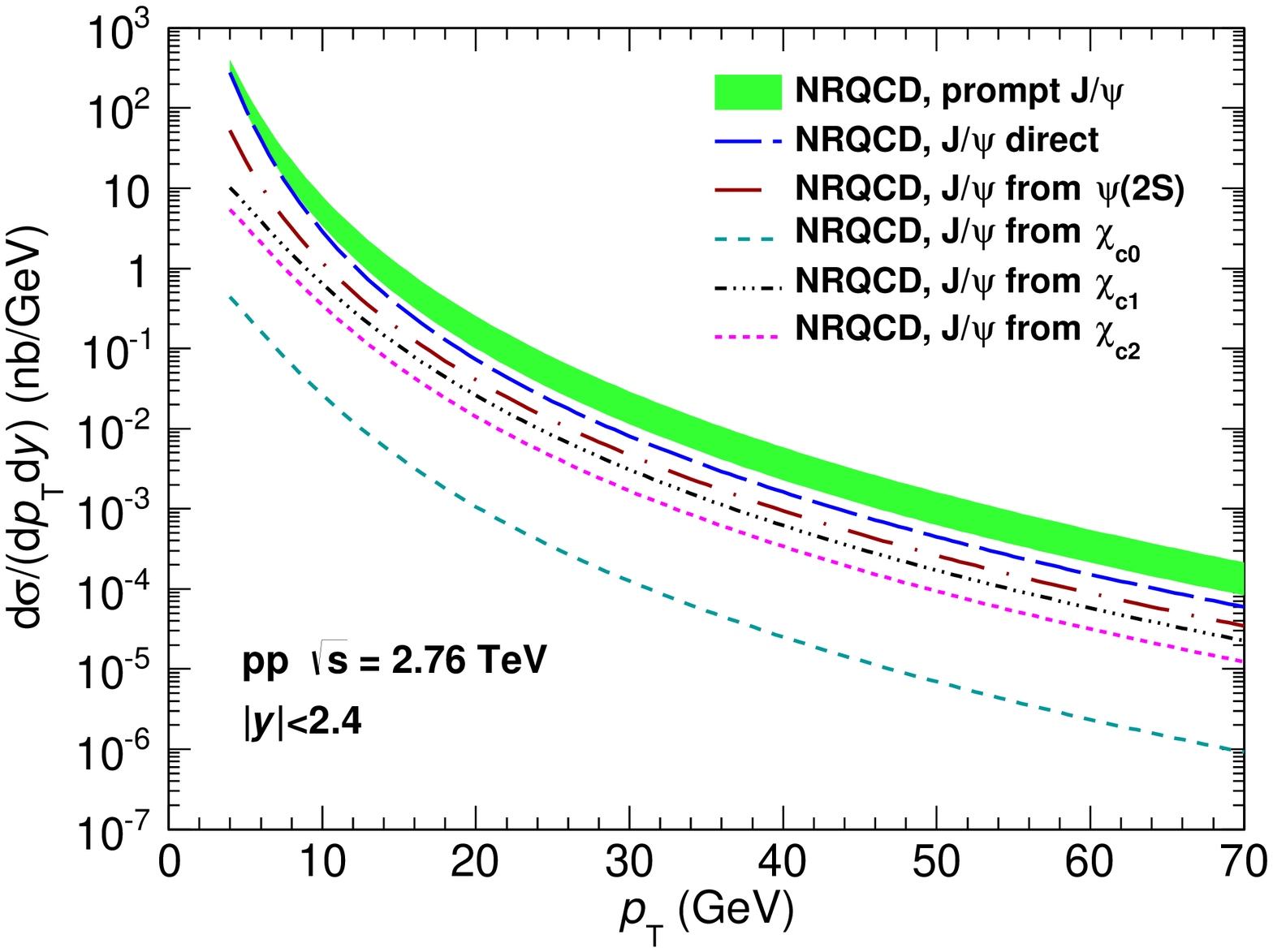}~~~~ \includegraphics[width=6.15cm,height=5.0cm,angle=0]{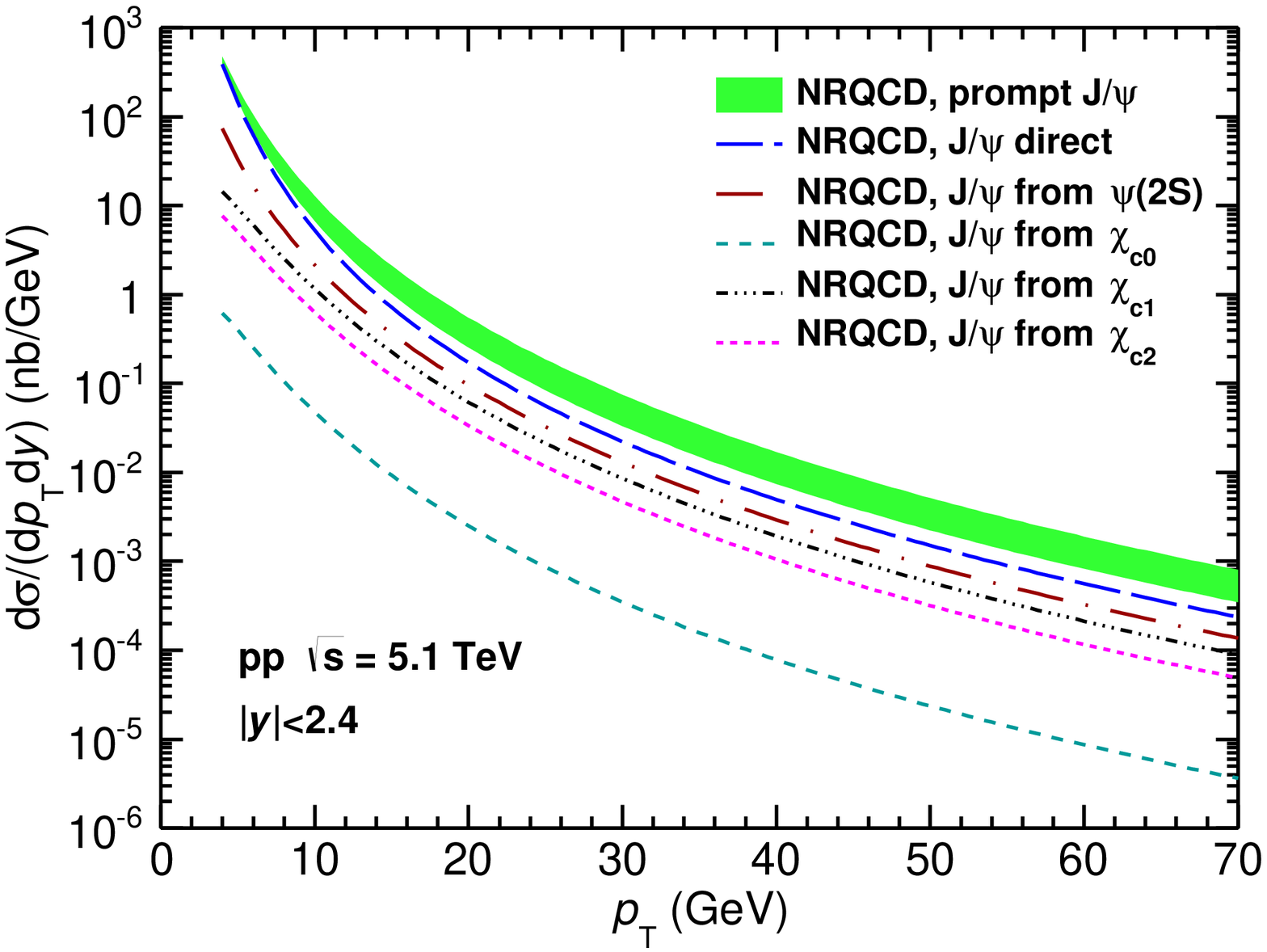}~~~~~\\ \includegraphics[width=6.15cm,height=5.0cm,angle=0]{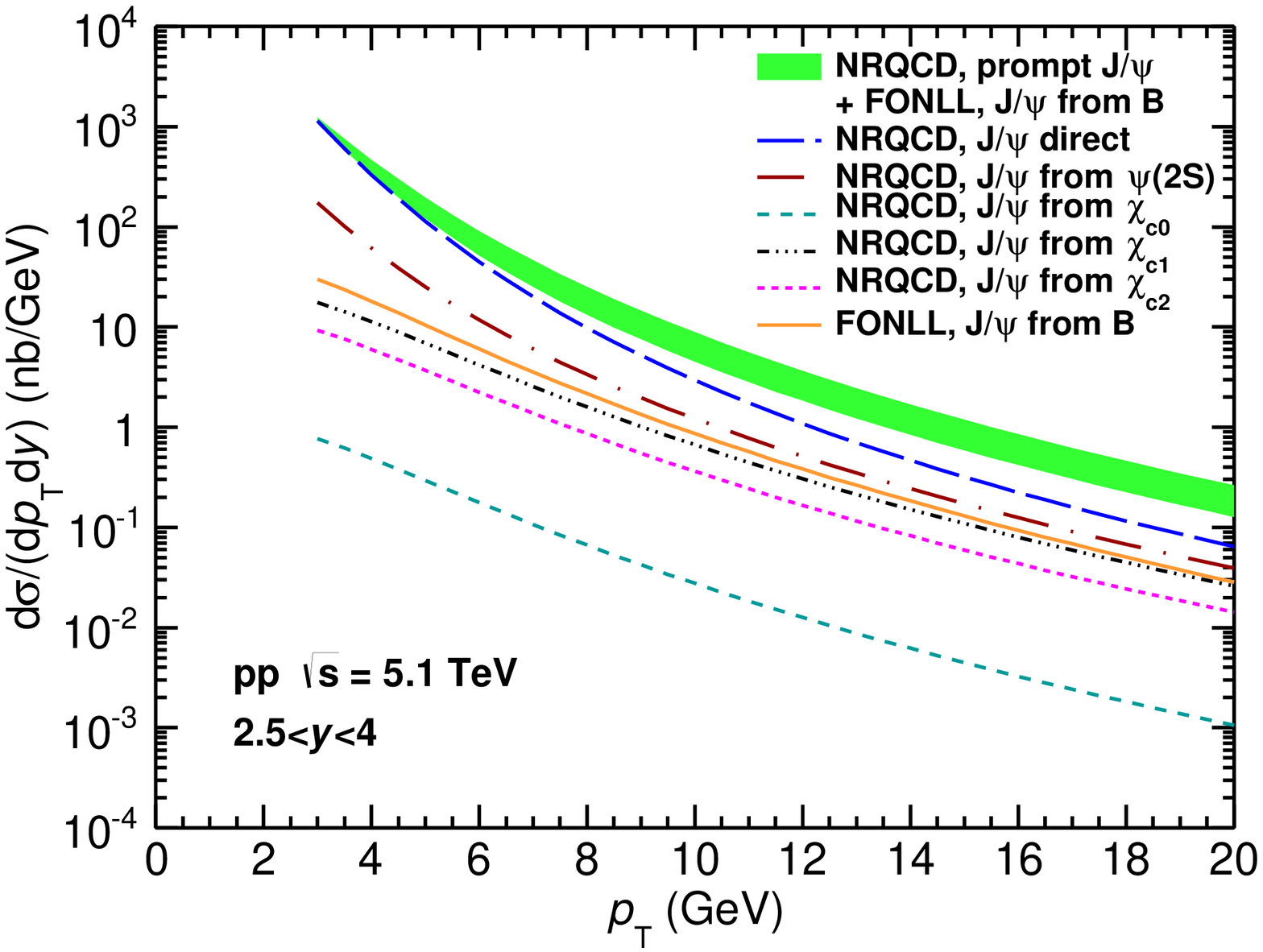}~~~~ \includegraphics[width=6.15cm,height=5.0cm,angle=0]{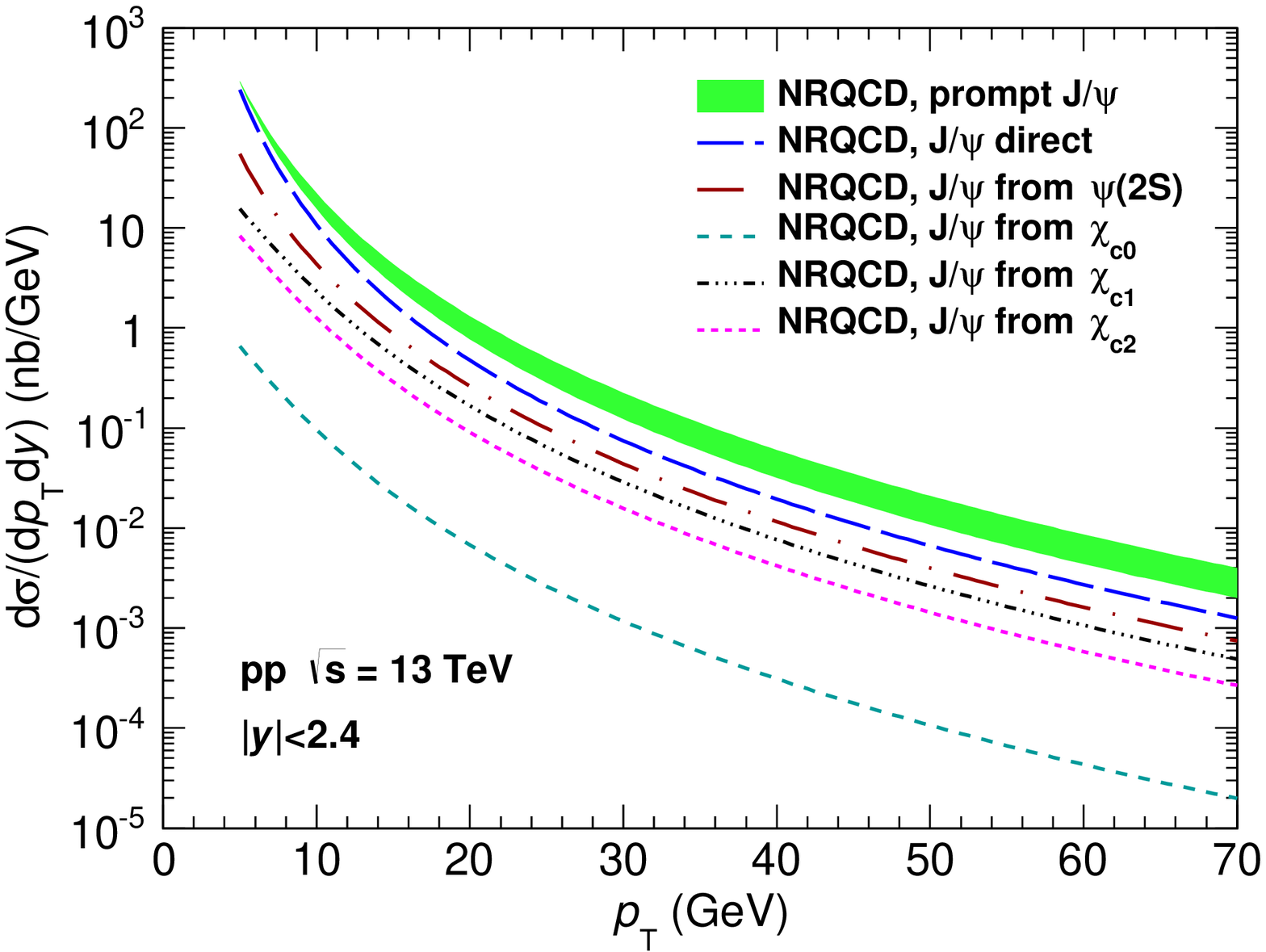}~~~~~\\ \includegraphics[width=6.15cm,height=5.0cm,angle=0]{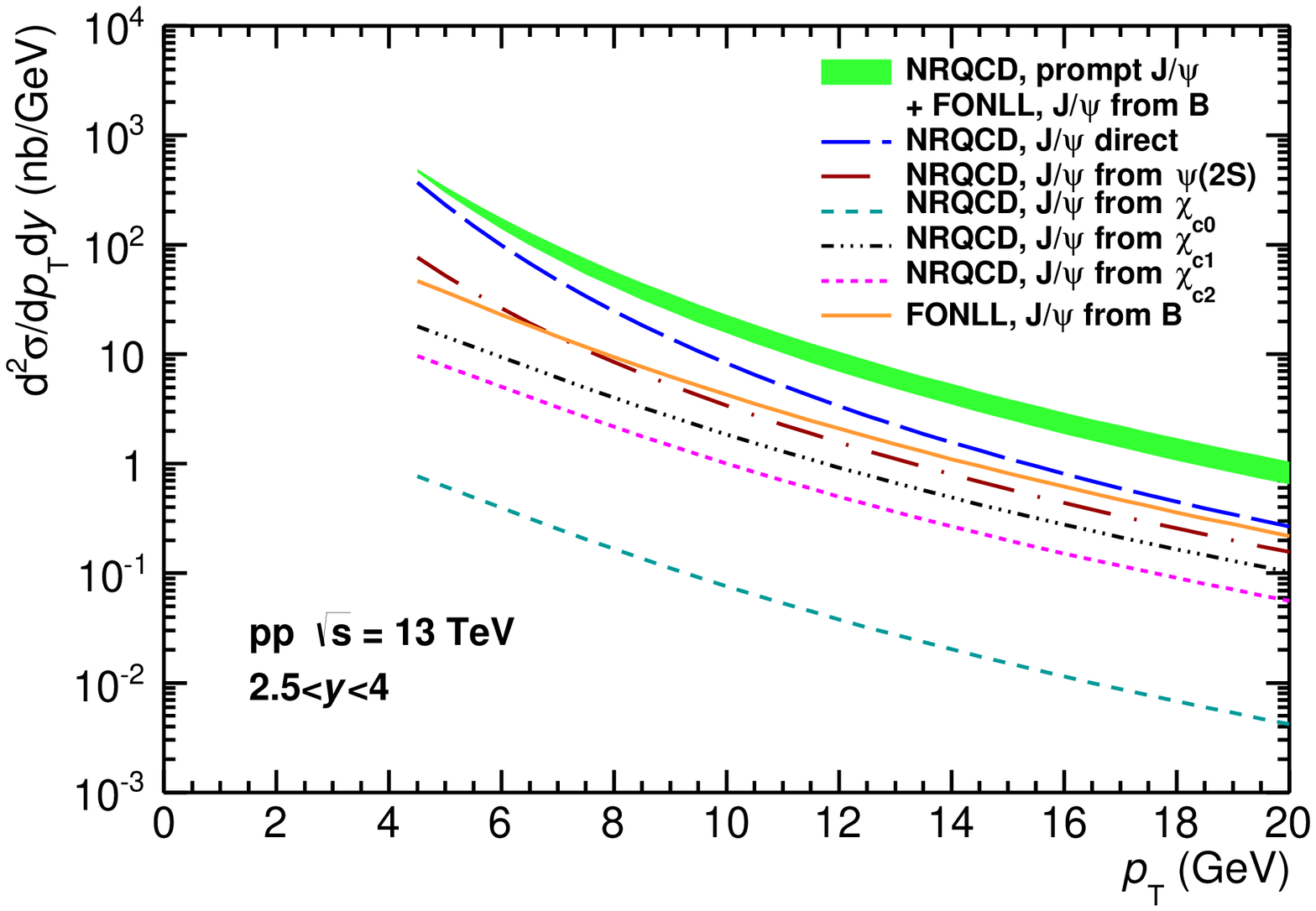} 
 \end{center}
 \caption{(Color online) Theoretical prediction for the differential cross-section of $J/\psi$ at $\sqrt{s}$ = 2.76, 5.1 and 13 TeV at mid and forward rapidity.}
 \label{fig5}
 \end{figure}

\begin{figure}[t]
 \begin{center}
 \includegraphics[width=6.15cm,height=5.0cm,angle=0]{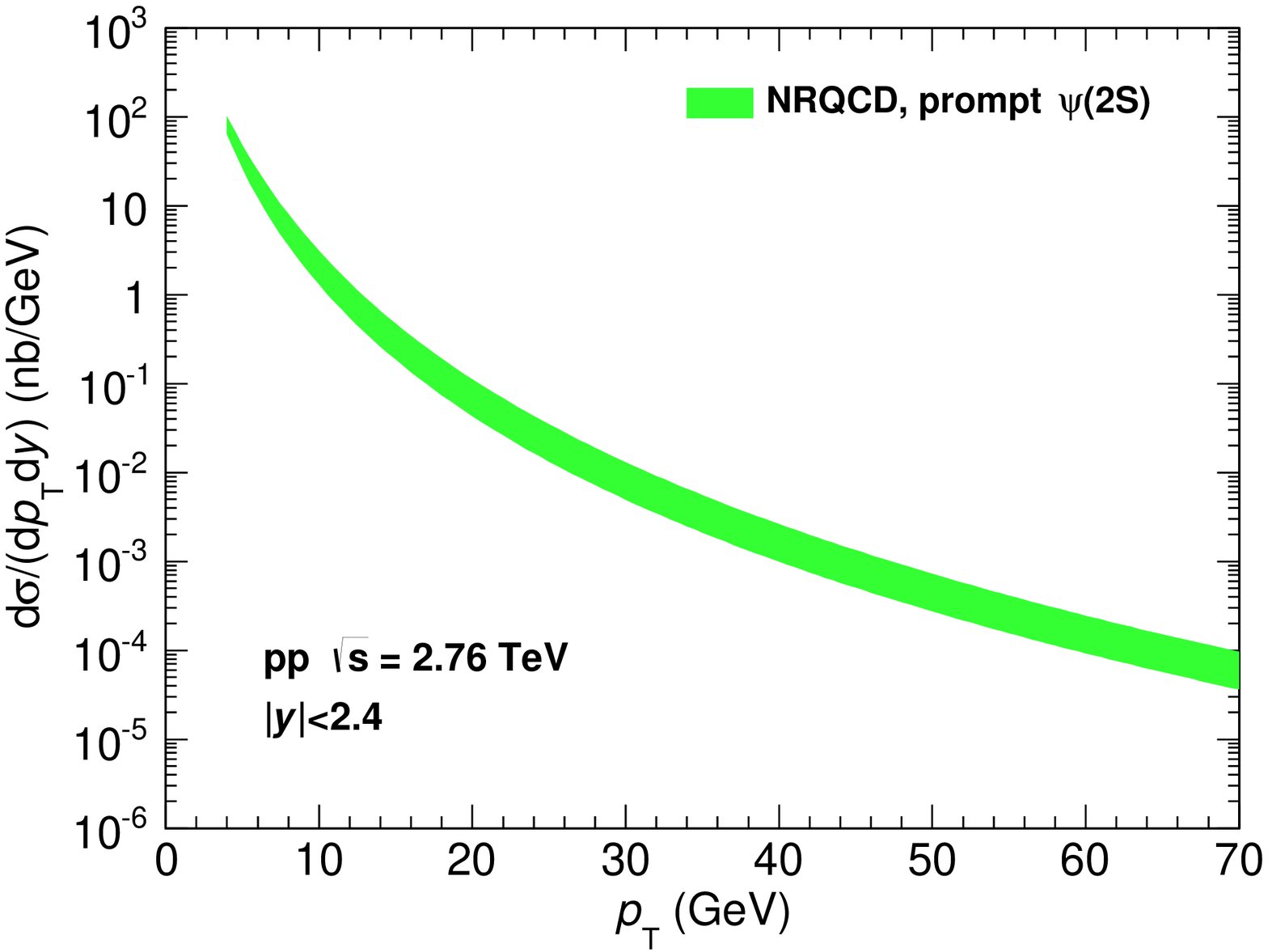}~~~~ \includegraphics[width=6.15cm,height=5.0cm,angle=0]{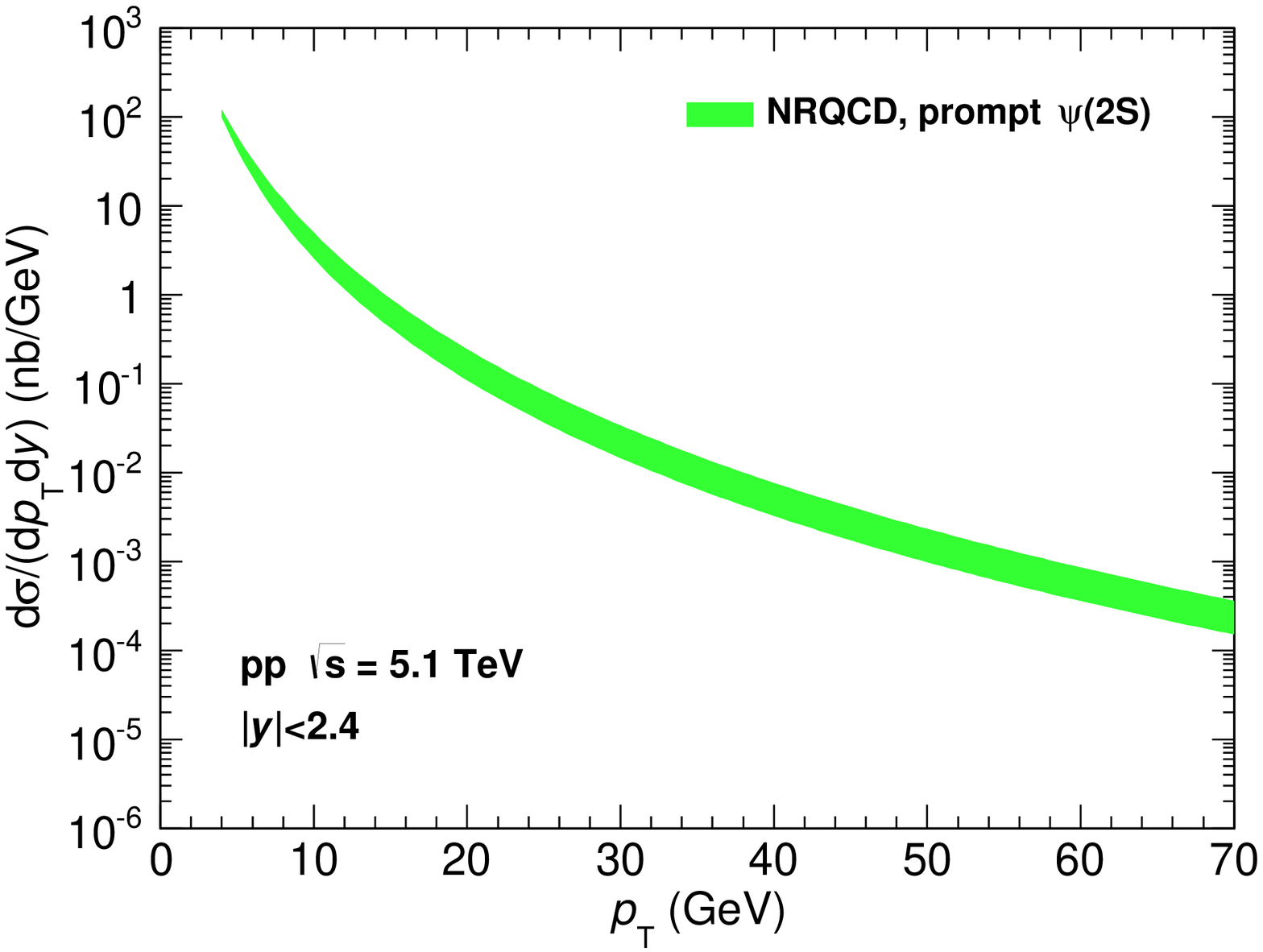}~~~~~\\  \includegraphics[width=6.15cm,height=5.0cm,angle=0]{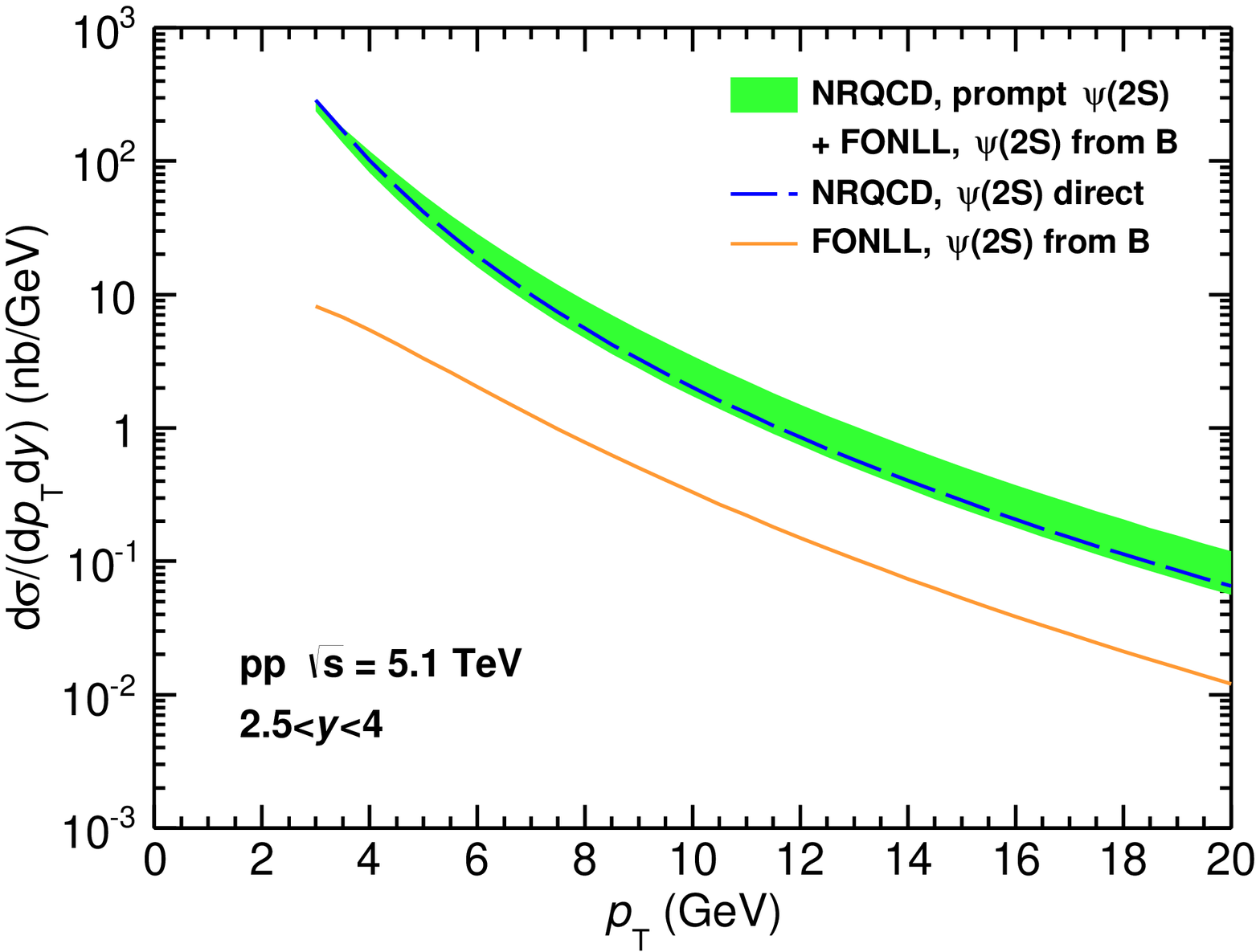}~~~~ \includegraphics[width=6.15cm,height=5.0cm,angle=0]{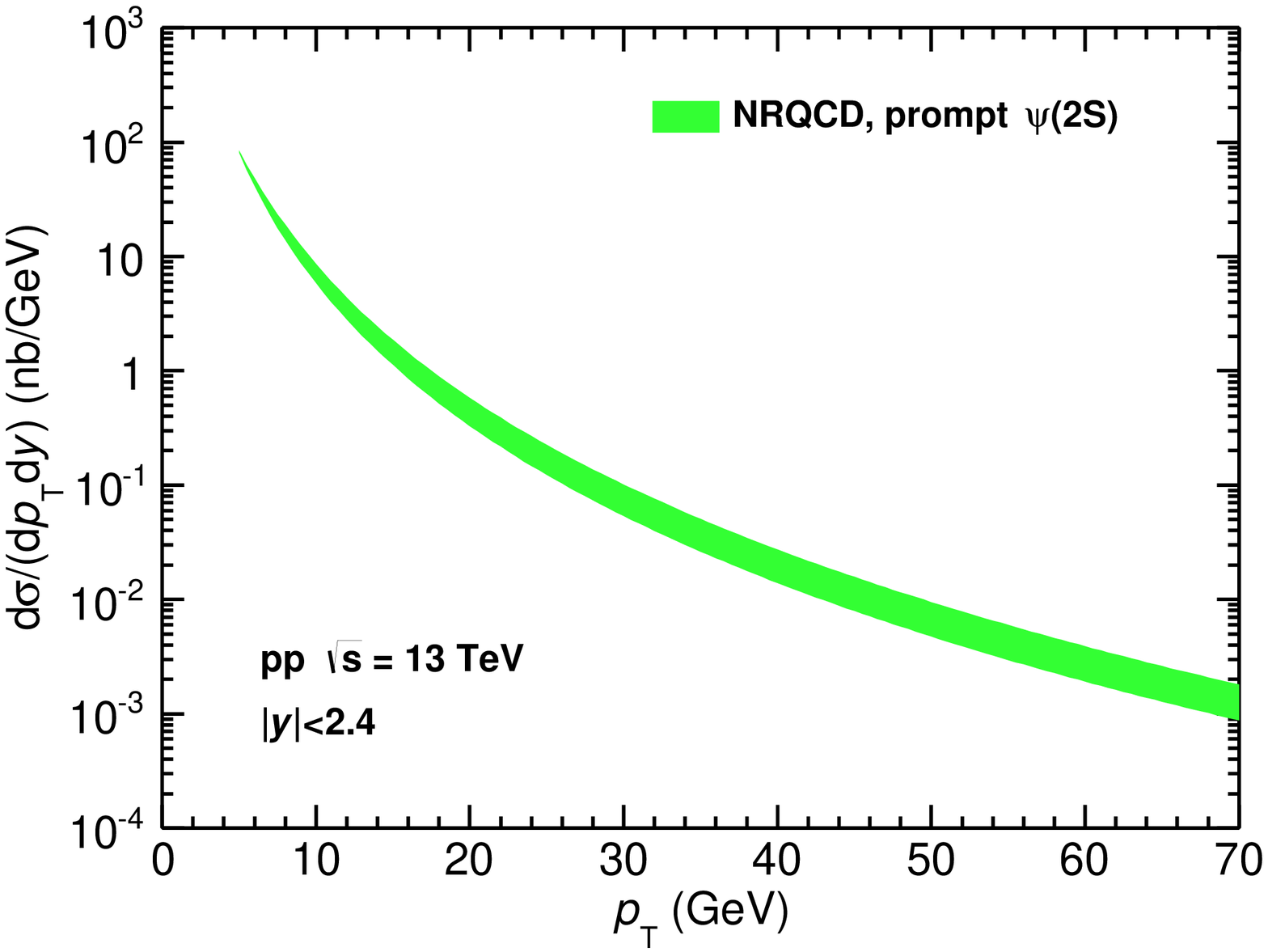}~~~~~\\ \includegraphics[width=6.15cm,height=5.0cm,angle=0]{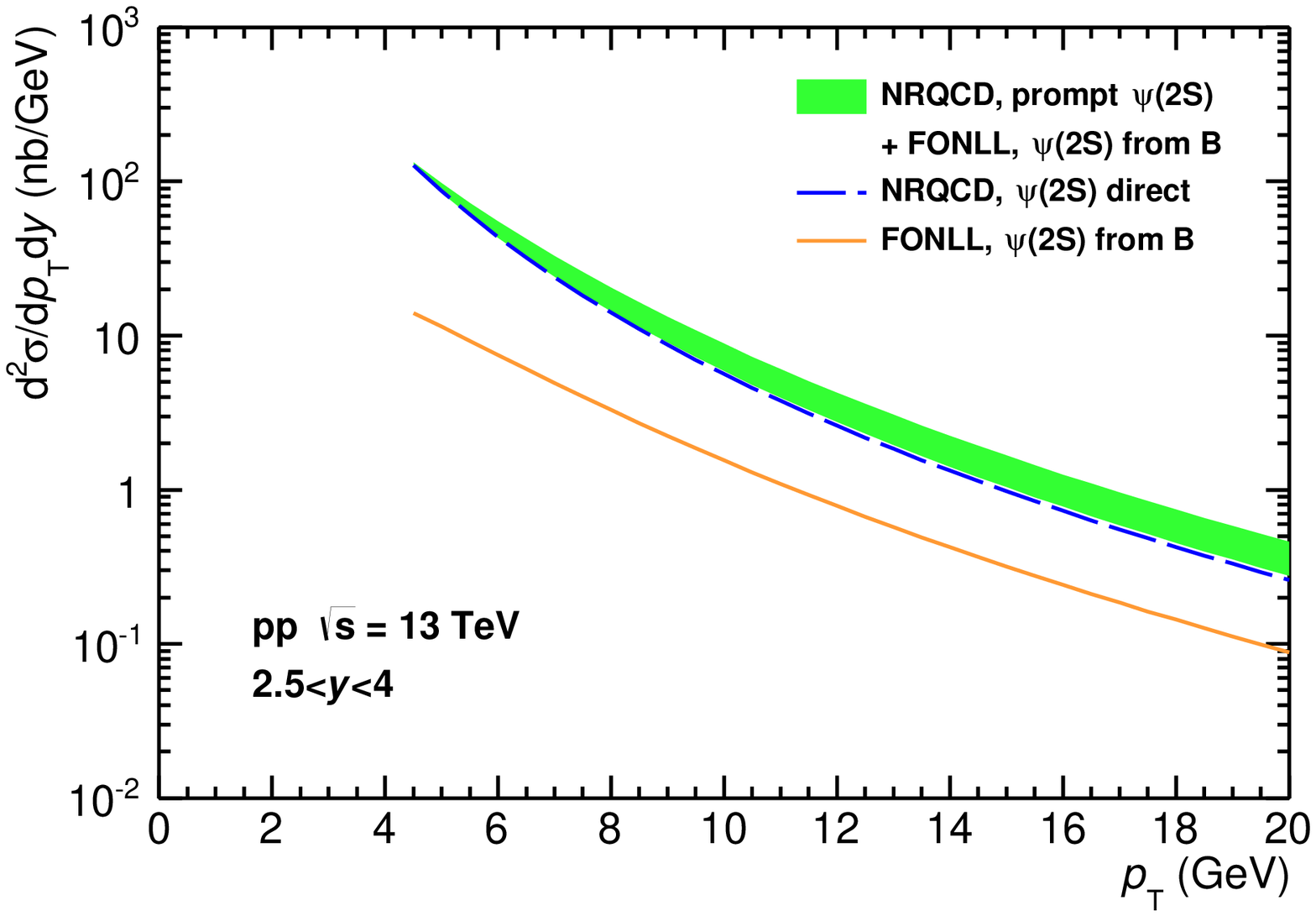}
 \end{center}
 \caption{(Color online) Theoretical prediction for the differential cross-section of $\psi(2S)$ at $\sqrt{s}$ = 2.76, 5.1 and 13 TeV at mid and forward rapidity.}
 \label{fig6}
 \end{figure}

This success of NRQCD calculations in describing the $p-p$ collisions data at various rapidities and energies, have prompted the predictions at $\sqrt{s}$ = 2.76, 5.1 and 13 TeV. These predictions have been shown in Fig.~\ref{fig5} and~\ref{fig6}. It will be interesting to test applicability of these calculations at much higher centre-of-mass energy of 13 TeV in 2015. On the other hand, the predictions at $\sqrt{s}$ = 2.76 and 5.1 TeV may be used for the normalization of the Pb-Pb collisions data.

\section{Summary and outlook}
In summary, the prompt and inclusive production cross-sections of $J/\psi$ and $\psi(2S)$ at LHC energies have been calculated within the framework of LO NRQCD and FONLL. These calculations include the contributions from direct production and from the decays of heavier charmonium states such as $\psi(2S)$, $\chi_{c0}$, $\chi_{c1}$ and $\chi_{c2}$. The feed-down to $J/\psi$ and $\psi(2S)$ from $B$ meson decays has been implemented using the FONLL calculation. The comparisons with experimental data from LHC at different energies and rapidity windows show the LO NRQCD calculations give a good description of the production cross-sections of $J/\psi$ and $\psi(2S)$ for $p_T >$ 4 GeV. The calculations for the prediction of production cross-sections of $J/\psi$ and $\psi(2S)$ at $\sqrt{s}$ = 2.76, 5.1 and 13 TeV has been carried out. 

It may be noted that the fragmentation process contributes to the charmonium production at high $p_T$~\cite{prl113} and inclusion of this process may further improve the calculations. The production cross-sections at low $p_T$ has been well reproduced with the CGC+NRQCD formalism~\cite{raju} which is important for ALICE and LHCb data. In future, we intend to adopt the CGC formalisms~\cite{raju,raju2} for quarkonium production in the low $p_{T}$ region to cover the entire $p_{T}$ range with the inclusion of all the feed-down contributions.

\section*{ACKNOWLEDGEMENTS}
It is a pleasure to thank Matteo Cacciari for helpful discussion about FONLL. The work of B. P. 
was supported by CSIR, India (File No. 09/489(0085)/2010-­EMR-­I).


\end{document}